\documentclass[12pt, a4paper]{article}

\usepackage{amsmath, amssymb, amsthm}
\usepackage{abstract}
\usepackage{color}
\usepackage{calc}
\usepackage{adjustbox}
\usepackage{url}
\usepackage{hyperref}

\usepackage{orcidlink}

\newtheorem{Corollary}{Corollary}
\newtheorem{Theorem}{Theorem}
\newtheorem*{Proof}{Proof}

\newtheorem{Definition}{Definition}
\newtheorem{Example}{Example}
\newtheorem{Problem}{Problem}

\begin{document}

{\LARGE\centering{\textbf{Shadow economy as a null set. Measure-theoretic approach to shadow economy}}}

\vspace{0.4cm}
\begin{center}
Rados\l{}aw A. Kycia $^{a}$ \orcidlink{0000-0002-6390-4627},
Andrzej Buszko $^{b}$ \orcidlink{0000-0003-0600-4646}
\end{center}

\begin{center}
$^a$ Institute of Mathematical Economics, Collegium of Economic Analysis, Warsaw School of Economics, Warsaw, Poland \\
$^b$ Pomeranian University in S\l{}upsk, S\l{}upsk, Poland
\end{center}

\begin{abstract}
\noindent
In this paper we asked if the shadow economy can be defined as a system of null sets for institutional measures or sensors that measure economic transactions. As an answer a measure-theoretic testable framework to redefine and quantify the shadow economy was designed. Traditional economic definitions frequently conflate the actual shadow economy with the observational limitations of state institutions. To resolve this, we model the shadow economy as a null set relative to institutional measures and analyze three complementary economic regimes. We explore in this context how standard mathematical tools can be given economic meaning. First, in the low-resolution regime, we apply Banach Lattices and Lebesgue decomposition of a measure to address transaction underreporting by aggregating incomplete institutional registries using a supremum fusion operator. Second, for hidden economic dimensions, we utilize Copula theory and Kullback-Leibler divergence optimization to reconstruct unobserved financial flows from technological and physical indicators. Third, we address audit scalability by modeling the shadow economy as a sparse vector, applying Compressed Sensing to detect high-volume sparse evasion networks through random, incoherent, and privacy-preserving sampling.
\end{abstract}


\section{Introduction}

There are many different and sometimes contradictory definitions of the shadow economy, which makes comparing diverse empirical studies difficult. Terms such as illegal production, illicit economy, gray zone, informal sector production, moonlight economy, non-observed economy, non-registered business activity, unreported economy, and underground economy typically cover only specific facets of shadow economic activity. Feige \cite{ShadowEconomy_Feige} defines the shadow economy as economic activity that is unreported and therefore not officially measured. However, Frey and Schneider in \cite{FreySchneider2000} reduce the concept primarily to unregistered production activities, which may be considered an overly narrow approach. Such a definition leaves unresolved the status of services, barter, and financial transactions, despite the fact that these areas often constitute important channels of shadow-economic activity. Similarly, Tanzi \cite{VitoTanzin1999} associates the shadow economy with production excluded from national accounts. Nevertheless, this definition raises a conceptual ambiguity: if an activity is recorded within national accounts but the corresponding tax obligations are evaded, should it be excluded from the shadow economy merely because it has been statistically recorded?

A further limitation can be identified in the definition proposed by Dell'Anno \cite{DellAnnoSolomon_2008}, who characterizes the shadow economy as underground production comprising both informal and illegal activities. The distinction between these categories is not entirely clear. Informal activities frequently operate outside the legal and regulatory framework, which suggests a substantial overlap with illegality. As a result, the boundary between informal and illegal activities remains insufficiently specified. Defining the shadow economy exclusively through tax evasion is also overly simplistic and analytically unsatisfactory. Not every case of tax evasion necessarily constitutes participation in the shadow economy, while many shadow-economic activities involve features that extend far beyond tax non-compliance. Consequently, a purely tax-based definition lacks the precision required for rigorous analysis and does not provide a clear framework for identifying the boundaries of the shadow economy.

Smith-Hillman in \cite{SmithHillman_2007} also proposed a definition of the shadow economy. Nevertheless, limiting the concept to activities omitted from official GDP estimates and undetected by tax authorities constitutes a significant simplification. The existence of the shadow economy is independent of its identification by public institutions. Failure to detect such activities does not eliminate their economic consequences; it merely renders their scale uncertain. This issue is of particular importance because the shadow economy may function as a sensitive indicator of institutional inefficiency, excessive regulation, or deficiencies in governance. If its magnitude remains unknown, its explanatory role can be incorrectly attributed to other economic variables, leading to flawed diagnoses and misguided policy recommendations. Therefore, definitions that equate the shadow economy with its statistical observability risk confusing the phenomenon itself with the capacity of state institutions to measure it. A unified definition of the shadow economy should refer to all economic activities that operate outside the legal or regulatory framework and/or generate adverse consequences for the public sector through the loss of budgetary revenues.

A rigorous approach should identify relevant ingredients, operationalize them using measurable indicators, assign weights theoretically or empirically, and integrate them within a multidimensional framework (e.g., MIMIC models, composite indices, structural equation models). Academics and researchers face persistent challenges with shadow economy measurement. The most widely recognized and employed tool is the MIMIC (Multiple Indicators Multiple Causes) model \cite{DelAnnoSchneider2005, DelAnno2007}. While its core concept is well motivated, the specific variables chosen for shadow economy assessment often vary across studies. For example, Dell'Anno and Schneider in \cite{DelAnnoSchneider2005} identify variables closely connected with shadow economy dynamics: tax burden, unemployment, self-employment, government consumption, an index of justice efficacy, and an index of illegality. However, in \cite{DelAnno2007}, Dell'Anno used the following determinants: government employment, tax burden, subsidies/GDP, social benefits paid by government, unemployment, and self-employment. Because different explanatory variables were applied, the resulting estimates cannot be directly compared. Whenever MIMIC is employed, another critical requirement is the verification of time-series stationarity, with particular emphasis on unit-root testing. Many publications overlook the necessity, in cases of non-stationarity, of data adjustment through logarithmic transformation or first-order differencing prior to parameter estimation and goodness-of-fit evaluation.

Given these methodological challenges, new mathematical approaches to modeling the shadow economy are needed to capture its full structural spectrum. We therefore propose an abstract, bird's-eye-view model based on measure theory \cite{Folland, Measure_Schilling, MeasureTheoryI_Bogachev} that isolates the structural causes of unrecorded activity across three distinct regimes. This structural information can be used to improve indirect economy measurements. As a consequence, suitable adjustments in government data engineering (GovTech \cite{GovTech}) allow authorities to monitor the shadow economy with higher precision.

While there is no identical measure-theoretic treatment in the literature, several modern approaches employ statistical, optimal transport, and machine learning methods. The MIMIC framework faces a fundamental scale-identification problem, making calibration central, and several calibration rules have been proposed to assign a unit of measure to latent scores \cite{DelAnno}. In stochastic modeling, hidden Markov models have been used to infer latent ensemble flows from aggregate records \cite{HiddenMarkovChains}. As an alternative, multi-marginal optimal transport formulations have been proposed for population and funding flows, aggregating observations as constrained marginals and encoding time-dependent transitions through learned cost functions \cite{OptimalTransport}. A distinct, non-stochastic but measure-like treatment of the tax gap reconstructs indirect tax accountability as agreement among physical, transactional, and fiscal records of the same goods, formalizing agreement through reconciliation residuals and noise floors of independent checks \cite{FundingMeasureLike}. Another direction models tax gap behavior and compliance using Random Forest and Gradient Boosting algorithms combined with econometric inference and high-frequency digital data such as mobile money flows and electronic invoicing \cite{BigDataApproach}.

The paper is organized as follows: Section \ref{Section_IntuitivePicture} provides an intuitive introduction to the measure-theoretic formulation. Sections \ref{Section_LowResolutionRegime}, \ref{Section_HiddenDimensionsRegime}, and \ref{Section_CS} present three complementary regimes of shadow economy modeling focused on transaction underreporting (Banach lattices and measure fusion), missing institutional measures along hidden dimensions (copulas and Kullback-Leibler divergence optimization), and scalable auditing under sparsity (Compressed Sensing), respectively. Finally, we synthesize the results and provide recommendations for decision-makers.

Since mathematical results used in the manuscript are well-established, we defer the full proofs to Supplementary materials.

\section{Intuitive picture}
\label{Section_IntuitivePicture}

We emphasize that our models describe a static snapshot of the shadow economy. Because measurement and enforcement actions alter agent behavior - as captured by Goodhart's law \cite{Goodhart_law} and the Lucas critique \cite{Lucas_law} - the static framework presented here serves as a foundational first approximation toward a full dynamic theory.

We will heavily rely on measure theory described, e.g., in \cite{Folland, Measure_Schilling, MeasureTheoryI_Bogachev}. The underlying measurable space $(\Omega, \mathcal{F})$, where $\Omega$ is a set of economic transactions and $\mathcal{F}$ its $\sigma$-algebra, represents the totality of economic transactions within a given jurisdiction (ranging from a municipality to a nation-state or the global economy). On this space, we distinguish two fundamental classes of measures:
\begin{itemize}
 \item Institutional measures, which are deployed by state agencies to observe, record, and tax economic activity. Because institutional registries are incomplete, they fail to capture the entirety of $\Omega$, giving rise to the shadow economy as a collection of institutional null sets or underreported sets.
 \item The true (full) measure, which captures the actual volume or monetary value of all economic transactions, including concealed ones. This measure is a latent construct that is not directly and fully observable by the government, but can be approximated or reconstructed using the mathematical regimes developed below.
\end{itemize}

Transitioning from a coarser institutional measure to a finer one corresponds to tightening the fiscal and regulatory monitoring approach. In this language, 'equality almost everywhere' of two economies with respect to an institutional measure means that they appear identical in official statistics while potentially differing  on institutional null sets (the shadow economy). If we let $\nu^*$ denote the true transaction-volume measure on $(\Omega, \mathcal{F})$ and $f: \Omega \to \mathbb{R}_+$ denote a measurable, $\nu^*$-integrable function representing the added value of a transaction $x \in \Omega$, then the Lebesgue integral $\mu^*(A) = \int_A f(x) \, d\nu^*(x)$ defines the true monetary value created within any measurable subset of transactions $A \in \mathcal{F}$.

Within this interpretational perspective, we identified components of the shadow economy that arise from four primary structural origins:
\begin{itemize}
 \item Methodological Shadow Economy - results from excluding specific classes of transactions from measurement by statutory or statistical design (e.g., unpaid household production, de minimis tax-exempt thresholds, or strictly prohibited illicit trade).
 \item Technological Shadow Economy - results from the absence of state sensors capable of recording specific transaction channels (e.g., peer-to-peer physical cash or privacy-shielded digital exchanges when only electronic banking registries are monitored).
 \item Topological (Juristical) Shadow Economy - results from transactions routed outside the territorial or legal jurisdiction of domestic institutions (e.g., offshore tax havens), which remain invisible unless covered by cross-border automatic information exchange treaties.
 \item Shadow Economy from Information Asymmetry - results from reliance on self-reported declarations in the absence of independent objective verification, allowing systematic underreporting of transaction values.
\end{itemize}
These contributions will be visible in the following sections.

Rather than modeling individual psychological propensities to evade taxes, our measure-theoretic approach formalizes institutional observability - the structural capacity of state registries to measure the economy. Under this lens, both statutory tax exemptions and illicit tax evasion correspond to sets of measure zero (or zeros of reduced Radon-Nikodym density) with respect to institutional measures.

\section{Low resolution regime}
\label{Section_LowResolutionRegime}
We start from the case where the zero sets are present due to low resolution of existing institutional measures. We will define the shadow economy as a null set and consider how to improve such measures to remove these blind spots.

Consider the measurable space $(\Omega, \mathcal{F})$ of all economic transactions. Let $\nu^*$ be a $\sigma$-finite true transaction-volume measure on $(\Omega, \mathcal{F})$, and let $f: \Omega \to \mathbb{R}_+$ be a measurable function representing the added value of a transaction $x \in \Omega$. Throughout this paper, we define the \textit{true value measure} $\mu^*$ on $(\Omega, \mathcal{F})$ by
\begin{equation}
  \mu^*(A) = \int_A f(x) \, d\nu^*(x), \quad A \in \mathcal{F},
\end{equation}
assuming $\mu^*(\Omega) < \infty$ (finite total value of the economy). Equivalently, whenever $\mu^*(A)$ denotes the true monetary volume of $A$, we may view $f(x)$ as the value density with respect to $\nu^*$ or normalize $f(x) \equiv 1$ when integrating directly with respect to $d\mu^*$.

We contrast $\mu^*$ with a finite family of \textit{institutional measures} $\{\mu_i\}_{i \in I}$, where $I$ is a finite index set representing distinct state registries (e.g., VAT declarations, corporate income tax filings, social security records, customs registries). Assuming that institutional registries do not influence nonexistent economic value on any measurable set, each registry satisfies $0 \le \mu_i(A) \le \mu^*(A)$ for all $A \in \mathcal{F}$.

To aggregate the institutional registries without double-counting transactions that appear simultaneously in multiple registries, we define the combined institutional measure $\mu$ as the lattice supremum, \cite{BanachLattice}, across $i \in I$:
\begin{equation}
  \mu = \bigvee_{i \in I} \mu_i.
  \label{Eq_TotalInstitutionalMeasure}
\end{equation}
This means that if the registries $\{\mu_i\}_{i \in I}$ are supported on pairwise disjoint economic domains, the supremum in (\ref{Eq_TotalInstitutionalMeasure}) reduces to the ordinary sum $\mu = \sum_{i \in I} \mu_i$. Below, we will focus on the structure of Banach space of measures, and then the supremum operation will be important extending the space to Banach lattice \cite{BanachLattice, MeasureTheoryI_Bogachev}.

As institutional measures only record activity present in the true economy ($0 \le \mu \le \mu^*$), $\mu$ is absolutely continuous \cite{Folland} with respect to the true measure $\mu^*$ that is denoted by
\begin{equation}
  \mu \ll \mu^*.
\end{equation}
Since $\mu^*$ is a finite positive measure and $\mu \ll \mu^*$ with $0 \le \mu \le \mu^*$, the Radon-Nikodym theorem \cite{Folland, Measure_Schilling} guarantees the existence of a unique density $g \in L^1(\Omega, \mathcal{F}, \mu^*)$ bounded in $[0,1]$:
\begin{Definition}[Transaction visibility]
The visibility of a transaction $x \in \Omega$ is defined as the Radon-Nikodym derivative
\begin{equation}
  g(x) = \frac{d\mu}{d\mu^*}(x) = \max_{i \in I} \frac{d\mu_i}{d\mu^*}(x) \in [0,1] \quad (\mu^*\text{-a.e.}).
\end{equation}
\end{Definition}
A transaction $x$ is fully visible to the state when $g(x)=1$, partially underreported when $0 < g(x) < 1$, and completely invisible when $g(x)=0$. This motivates distinguishing the measurable support of the completely hidden shadow economy from the family of shadow-economy null sets in the following
\begin{Definition}[Shadow economy support and null family]
The \textit{shadow economy support} (defined up to $\mu^*$-null sets) is the measurable set of completely unrecorded transactions:
\begin{equation}
  S = \{x \in \Omega \mid g(x) = 0\} \in \mathcal{F},
\end{equation}
and the \textit{family of shadow economy null sets} is
\begin{equation}
  \mathcal{S} = \left\{A \in \mathcal{F} \;\middle|\; \mu(A) = \int_A g(x)\, d\mu^*(x) = 0 \text{ and } \mu^*(A) > 0\right\}.
\end{equation}
\end{Definition}

Null sets contributing to the shadow economy arise from distinct legal and technical mechanisms mentioned above that are distinguished in the following definitions.
\begin{Definition}[Threshold shadow economy]
For a statutory turnover threshold $C > 0$, the measurable set
\begin{equation}
  S_{\text{threshold}} = \{x \in \Omega \mid f(x) < C\}
\end{equation}
is a null set for the thresholded institutional measure
\begin{equation}
  \mu_i(A) = \int_A \chi_{\{f(x) \ge C\}}(x) \, d\mu^*(x).
\end{equation}
\end{Definition}
For example, small-scale unregistered sole proprietorships or hobbyist sales legally exempt from registration below a statutory revenue cap $C$ fall into $S_{\text{threshold}}$.

\begin{Definition}[Technological shadow economy]
An institutional measure $\mu_i$ exhibits a technological 'blind spot' on a measurable set $S_{\text{tech}, i} \in \mathcal{F}$ with $\mu^*(S_{\text{tech}, i}) > 0$ if $\mu_i(S_{\text{tech}, i}) = 0$ due to the inability of institutional sensor $i$ to observe that transaction modality (e.g., physical cash transactions when $\mu_i$ monitors only electronic wire transfers).
\end{Definition}

We also characterize the core shadow economy that escapes every institutional registry simultaneously.
\begin{Definition}[Combined shadow economy support]
The combined shadow economy support invisible to all institutional registries $i \in I$ is the measurable set
\begin{equation}
\label{Def_ShadowEconomyCombined}
  S_c = \bigcap_{i \in I} \left\{x \in \Omega \;\middle|\; \frac{d\mu_i}{d\mu^*}(x) = 0\right\},
\end{equation}
and the corresponding collection of common null sets is $\mathcal{S}_c = \bigcap_{i \in I} \{A \in \mathcal{F} \mid \mu_i(A) = 0\}$.
\end{Definition}
Since each $\mu_i$ is a non-negative measure and $\mu = \bigvee_{i \in I} \mu_i$, we have $\mu(S_c) = 0$.

We now formulate the optimal fiscal enforcement problem on $[0,1]$.
\begin{Theorem}[Optimal condition for institutional measure]
Suppose the operational cost $c: \Omega \times [0,1] \to \mathbb{R}_+$ of enforcing visibility $g \in [0,1]$ for transaction $x \in \Omega$ is a Carath\'eodory function (measurable in $x$ for each $g \in [0,1]$, and convex and continuously differentiable with respect to $g \in [0,1]$ for $\nu^*$-almost every $x \in \Omega$) satisfying $\sup_{g \in [0,1]} c(x,g) \in L^1(\Omega, \mathcal{F}, \nu^*)$. Given a tax rate $\tau \in [0,1]$ and $\nu^*$-integrable added-value function $f(x)$, a measurable visibility profile $g^*(x) \in [0,1]$ solves the net budget revenue maximization problem
\begin{equation}
  \max_{g: \Omega \to [0,1]} \int_{\Omega} \left( \tau f(x) g(x) - c(x, g(x)) \right) d\nu^*(x)
  \label{Eq.OptimalMeasure}
\end{equation}
if and only if for $\nu^*$-almost every $x \in \Omega$ we have
\begin{equation}
  \begin{cases}
    \tau f(x) \le \frac{\partial c}{\partial g}(x, 0) & \text{if } g^*(x) = 0, \\[4pt]
    \tau f(x) = \frac{\partial c}{\partial g}(x, g^*(x)) & \text{if } g^*(x) \in (0, 1), \\[4pt]
    \tau f(x) \ge \frac{\partial c}{\partial g}(x, 1) & \text{if } g^*(x) = 1.
  \end{cases}
  \label{Eq_KKT_Visibility}
\end{equation}
Here $\frac{\partial c}{\partial g}(x, g(x))$ represents the marginal enforcement cost of raising transaction visibility at $x$.
\end{Theorem}
\begin{Proof}
The proof is an application of Theorem 18.19 from \cite{InfiniteDimensionalAnalysis}.

Since $c(x,g)$ is a Carath\'eodory function with an integrable majorant and the integrand in (\ref{Eq.OptimalMeasure}) depends pointwise on $g(x) \in [0,1]$, it guarantees that maximizing the integral over measurable functions $g: \Omega \to [0,1]$ reduces (see also Theorem 14.60 of \cite{VariationalAnalysis_RockafellarWets}) to maximizing the pointwise objective function $I_x(g) = \tau f(x) g - c(x, g)$ over the compact interval $g \in [0,1]$ for $\nu^*$-almost every $x \in \Omega$.

Since $g \mapsto c(x, g)$ is convex and continuously differentiable on $[0,1]$, $I_x(g)$ is concave and continuously differentiable on $[0,1]$. By the Kuhn-Tucker optimality conditions on $[0,1]$, $g^* \in [0,1]$ maximizes $I_x(g)$ if and only if $I_x'(0) \le 0$ when $g^* = 0$, $I_x'(g^*) = 0$ when $g^* \in (0,1)$, and $I_x'(1) \ge 0$ when $g^* = 1$, which yields (\ref{Eq_KKT_Visibility}).
\end{Proof}
\begin{Corollary}
If $c(x,g)$ is convex with respect to $g \in [0,1]$, then for every transaction $x \in \Omega$ satisfying $\tau f(x) < \frac{\partial c}{\partial g}(x, 0)$, the unique optimal visibility in (\ref{Eq.OptimalMeasure}) is $g^*(x) = 0$.
\end{Corollary}
\begin{Proof}
For fixed $x$, define $I_x(g) = \tau f(x) g - c(x,g)$. Then $I_x'(g) = \tau f(x) - \frac{\partial c}{\partial g}(x,g)$. By convexity of $c(x, \cdot)$, the marginal cost $\frac{\partial c}{\partial g}(x,g)$ is non-decreasing in $g \in [0,1]$, so $\frac{\partial c}{\partial g}(x,0) \le \frac{\partial c}{\partial g}(x,g)$ for all $g \in [0,1]$. Since $\tau f(x) < \frac{\partial c}{\partial g}(x,0)$, we have $I_x'(g) < 0$ for all $g \in [0,1]$. Thus $I_x(g)$ is strictly decreasing on $[0,1]$ and attains its unique maximum at the boundary $g^*(x) = 0$.
\end{Proof}

This corollary formalizes why rational tax administrations deliberately leave micro-transactions in a legal shadow economy ($g^*(x)=0$): whenever the marginal administrative cost of monitoring exceeds the potential tax yield, enforcing visibility reduces net public revenue. Only in the idealized limit $c \equiv 0$ is it optimal for $\mu$ to fully recover $\mu^*$. The topic from economic perspective is recognized. Economic discussions of optimal tax administration are provided in \cite{OptimalTaxSystem, OptimaltaxSystem2}, and \cite{StateCapacity} discusses state capacity.

We now provide a simple example that explains the various null sets of shadow economy.
\begin{Example}
Consider the agricultural fruit production sector, where $\Omega$ represents all transactions and $f(x)$ is the monetary value of transaction $x$ in local currency. We distinguish three subsets of $\Omega$:
\begin{itemize}
 \item Organized retail and wholesale trade $A \subset \Omega$, where $g(x) = 1$ for $x \in A$. Transactions are fully invoiced and visible in electronic registries.
 \item Small-scale street and farm-gate trading $B \subset \Omega$, where transaction values $f(x)$ are small and $\tau f(x) < \frac{\partial c}{\partial g}(x, 0)$. The state sets $g(x) = 0$ on $B = S_{\text{threshold}}$ via a statutory exemption threshold.
 \item Off-the-books wholesale cash transactions $C \subset \Omega$, where $g(x) = 0$ for $x \in C$ due to technological blind spots in cash monitoring, even though $\tau f(x) > \frac{\partial c}{\partial g}(x, g(x))$ would make taxation profitable under improved sensing.
\end{itemize}
In this example, the total unrecorded shadow economy support is $S = B \cup C$.
\end{Example}

Using this framework, we (re)define four quantitative shadow-economy metrics. Some of them are already used in modern economical analyses and we provide measure-theoretic foundations.
\begin{Definition}[Macro gap]
Defined as
  \begin{eqnarray}
   \Delta := \int_{\Omega} f(x) \, d\nu^*(x) - \int_{\Omega} f(x) g(x) \, d\nu^*(x) = \notag \\
   =\int_{\Omega} (1 - g(x)) \, d\mu^*(x) = \mu^*(\Omega) - \mu(\Omega).
  \end{eqnarray}
  When $\mu^*$ can be estimated by physical traces such as electricity consumption \cite{ElectricityShadow1, ElectricityShadow2}, we can compute the relative macro gap
  \begin{equation}
   M_{\text{gap}} = \frac{\int_{\Omega} (1 - g(x)) \, d\mu^*(x)}{\mu^*(\Omega)} \in [0,1].
  \end{equation}
  Here $1 - g(x)$ describes the 'local invisibility density' of the economy, complementing classic input-output \cite{Leontief} and latent-variable MIMIC approaches \cite{Schneider_MIMIC1}.
\end{Definition}

\begin{Definition}[Bunching asymmetry metric]
Suppose that there exists some threshold, $C > 0$, which is a regulatory or tax-rate kink point (where the marginal tax rate or compliance cost jumps upward at $f(x) = C$, while an institutional registry $\mu$ continues to record declared values both below and above $C$). Taxpayers who manipulate or underreport transactions to remain just below $C$ create an excess transaction mass below $C$ and a missing mass above $C$, see \cite{BunchingPoint, Bunching} for usage of these measures defined in statistical terms. For $\epsilon > 0$, we define the bunching ratio
  \begin{equation}
   \mathcal{B}(C, \epsilon) = \frac{\mu\big(f^{-1}([C-\epsilon, C))\big)}{\mu\big(f^{-1}([C, C+\epsilon))\big)}.
  \end{equation}
  Here $\nu = \mu \circ f^{-1}$ is the pushforward value distribution on the Borel $\sigma$-algebra $\mathcal{B}(\mathbb{R}_+)$. Sharp bunching at the threshold manifests as an irregularity in the density of $\nu$ (or an atomic point mass at $C$) relative to the smooth counterfactual pushforward $\mu^* \circ f^{-1}$ when using true measure of transactions.
\end{Definition}

\begin{Definition}[Orthogonal measures and mismatch ratio]
Given two distinct institutional registries $\mu_1$ and $\mu_2$, we can take the Lebesgue decomposition \cite{Folland} of $\mu_1$ with respect to $\mu_2$
  \begin{equation}
   \mu_1 = \mu_1^{ac} + \mu_1^{\perp},
  \end{equation}
  where $\mu_1^{ac} \ll \mu_2$ (since every $\mu_2$-null set is also a $\mu_1^{ac}$-null set) and $\mu_1^{\perp} \perp \mu_2$ ($\mu_1^{\perp}$ - the singular part with respect to $\mu_{2}$, \cite{Folland}, is concentrated on a measurable set where $\mu_2$ vanishes). For any measurable domain $A \in \mathcal{F}$ with $\mu_1(A) > 0$, we define the normalized mismatch ratio
  \begin{equation}
   \mathcal{M}_{1,2}(A) = \frac{\mu_1^{\perp}(A)}{\mu_1(A)} = 1 - \frac{\mu_1^{ac}(A)}{\mu_1(A)} \in [0,1].
  \end{equation}
  This ratio quantifies the exact fraction of economic activity in $A$ detected by registry $\mu_1$ that is completely invisible to registry $\mu_2$.
\end{Definition}

\begin{Definition}[Marginal enforcement ROI]
Let $\mathcal{M} = \{\mu \mid \mu \ll \mu^*, \, 0 \le \mu \le \mu^*\}$ be the space of admissible institutional measures. The gross tax revenue functional is $R(\mu) = \int_{\Omega} \tau g(x) \, d\mu^*(x)$ and the administrative cost functional is $C(\mu) = \int_{\Omega} c(x, g(x)) \, d\nu^*(x)$. Under a nonnegative variation $\delta g(x) \ge 0$ supported on $A \in \mathcal{F}$, the first variations are
  \begin{eqnarray}
   \delta R(A) = \int_A \tau f(x) \delta g(x) \, d\nu^*(x), \\
   \delta C(A) = \int_A \frac{\partial c}{\partial g}(x, g(x)) \delta g(x) d\nu^*(x).
  \end{eqnarray}
  The marginal enforcement return on investment on $A$ is
  \begin{equation}
   \text{ROI}(A) = \frac{\delta R(A)}{\delta C(A)}.
  \end{equation}
  Whenever $\text{ROI}(A) < 1$, expanding institutional surveillance in domain $A$ costs more than the incremental tax revenue recovered \cite{Mirreless1, Mirreless2}.

\end{Definition}

\subsection{Estimation of total measure}

Because the true measure $\mu^*$ is unobservable, applied economics relies on indirect proxies and statistical estimators to approximate $\mu^*$ on macroscopic aggregates. We collect examples in the following list:
\begin{itemize}
 \item Electricity consumption: Comparing electric power demand to officially reported output reveals unrecorded production \cite{ElectricityShadow1, ElectricityShadow2}.
 \item Satellite night-light photometry: Remote sensing of nighttime luminosity $\mu_{\text{photo}}$ \cite{SateliteObservations} yields a first-order approximation $\mu^* \approx k \, \mu_{\text{photo}}$ via a calibrated luminosity-to-output conversion factor $k$.
 \item Currency demand: Excess demand for physical cash relative to cashless payment baselines estimates the volume of unrecorded transactions \cite{CashDemand1, CashDemand2}.
 \item Latent variable and state-space models: MIMIC structural equation models \cite{Schneider1, Schneider2, Schneider3, MIMIC1} and Kalman filtering of autoregressive unobserved components \cite{KalmanFilter, KalmanFilter_theory} generate macroscopic estimates $\mu_L \approx \mu^*$.
 \item Agent-Based Modeling (ABM): Monte Carlo simulations of synthetic economies $\Omega_{\text{sim}}$ \cite{ABS_ShadowModeling} allow exact tracking of simulated evasion $g_{\text{sim}} = \frac{d\mu_{\text{sim}}}{d\mu^*_{\text{sim}}}$ to calibrate real-world priors.
 \item Leontief Input-Output model \cite{Leontief, Leontief2} imbalance:  Given observed raw material inputs $\mu_{\text{in}}$ and a known technological output-to-input coefficient $k$ for a production sector, true output is approximated by $\mu^* \approx k \, \mu_{\text{in}}$, and the unrecorded shadow output is estimated by the residual $k \, \mu_{\text{in}} - \mu_{\text{out}}$ relative to officially reported output $\mu_{\text{out}}$.
\end{itemize}

We now formalize the fundamental trade-off between macroscopic consistency and microscopic non-identifiability when estimating $\mu^*$ from noisy aggregate of physical traces of economic activity.
\begin{Theorem}[Convergence of Coarse Measures]
\label{Th_ConvergenceOfCoarseMeasures}
Let $(\Omega, \mathcal{F})$ be the measurable space of economic transactions, and let $\mathcal{G} \subsetneq \mathcal{F}$ be a coarse sub-$\sigma$-algebra generated by macroeconomic aggregates (sectors). Let $(\Theta, \Sigma, P)$ be an external probability space on which the true measure $\mu^*(\theta, \cdot)$ is drawn from a Bayesian prior distribution over finite positive measures on $(\Omega, \mathcal{F})$, and let $\mu \ll \mu^*$ be an institutional measure.

For each macroscopic sector $A \in \mathcal{G}$, suppose we observe a sequence of noisy macroscopic physical traces
\begin{equation}
 Z_i(A) = T(\mu^*(A)) + \epsilon_i(A), \quad i = 1, \dots, n,
\end{equation}
where $T: \mathbb{R}_+ \to \mathbb{R}$ is a continuous bijection with a continuous inverse $T^{-1}$, and $\{\epsilon_i(A)\}_{i=1}^\infty$ are pairwise uncorrelated random variables on $(\Theta, \Sigma, P)$ independent of $\mu^*$, with $E[\epsilon_i(A)] = 0$ and $\text{Var}(\epsilon_i(A)) = \sigma^2 \in \mathbb{R}_+$.

Define the macroscopic empirical estimator on $A \in \mathcal{G}$ by
\begin{equation}
 \hat{\mu}^*_n(A) := T^{-1}\left(\frac{1}{n}\sum_{i=1}^n Z_i(A)\right).
\end{equation}
Let
\begin{equation}
 \mathcal{H}_n = \sigma(\{Z_i(A) : A \in \mathcal{G}, \, 1 \le i \le n\}) \subset \Sigma
\end{equation}
denote the $\sigma$-algebra generated by all macroscopic observations up to sample size $n$, and let
\begin{equation}
 \mathcal{H}_\infty = \sigma(\{\mu^*(A) : A \in \mathcal{G}\}) \subset \Sigma.
\end{equation}
Then the following conditions hold:
\begin{enumerate}
 \item For any microeconomic subset $B \in \mathcal{F} \setminus \mathcal{G}$ such that $\mu^*(B) \in L^2(\Theta, \Sigma, P)$ has non-degenerate conditional prior variance given the macroscopic aggregates, i.e., $c := E[\text{Var}(\mu^*(B) \mid \mathcal{H}_\infty)] > 0$, any sequence of estimators $\tilde{\mu}_n(B)$ measurable with respect to the macroscopic observations $\mathcal{H}_n$ has mean squared error bounded away from zero, i.e.,
 \begin{equation}
  \inf_{n \in \mathbb{N}} E\left[ \left(\tilde{\mu}_n(B) - \mu^*(B)\right)^2 \right] \ge c > 0.
 \end{equation}
 \item Conditional on any realization of $\mu^*$, for every macroscopic aggregate $A \in \mathcal{G}$, the estimator $\hat{\mu}^*_n(A)$ converges in probability to the true sector value $\mu^*(A)$ as $n \to \infty$, namely,
 \begin{equation}
  \forall \epsilon > 0, \quad \lim_{n \to \infty} P\left( \left\vert \hat{\mu}^*_n(A) - \mu^*(A) \right\vert > \epsilon \;\middle|\; \mu^*(A) \right) = 0.
 \end{equation}
 Consequently, the macroeconomic shadow-economy estimator $\hat{\Delta}_n(A) = \hat{\mu}^*_n(A) - \mu(A)$ is consistent on $\mathcal{G}$, i.e.,
 \begin{equation}
  \text{plim}_{n \to \infty} \hat{\Delta}_n(A) = \Delta(A).
 \end{equation}
\end{enumerate}
\end{Theorem}
\begin{Proof}
The proof is based on Bayesian inference on the space of parametric measures and noises $(\Theta, \Sigma, P)$. The measure in this sense is a random variable on this space, which allows us to use classical results of Hilbert space projection.

The first point is established by noting that the conditional expectation $\hat{\mu}_{\text{opt}}(B) = E[\mu^*(B) \mid \mathcal{H}_\infty]$ is the $L^2(P)$-optimal projection, see Section 9 of \cite{ProbabilityWithMartingales} or Section 27 of \cite{Measure_Schilling}. By the law of total variance, the conditional variance $E[\text{Var}(\mu^*(B) \mid \mathcal{H}_\infty)] = c > 0$ provides a strict lower bound on the mean squared error for any macroscopic estimator, preventing pointwise convergence on $\mathcal{F} \setminus \mathcal{G}$.

The second point follows directly from the Weak Law of Large Numbers applied to the observation noise and the Continuous Mapping Theorem for continuous inverse $T^{-1}$ \cite{Folland}.
\end{Proof}

Theorem \ref{Th_ConvergenceOfCoarseMeasures} provides three recommendations for fiscal monitoring:
\begin{itemize}
 \item When state institutions only observe macroscopic physical proxies on a coarse sub-$\sigma$-algebra $\mathcal{G}$, they cannot invert $\mu = g\mu^*$ for an individual firm $B \in \mathcal{F} \setminus \mathcal{G}$; however, at the sector level $A \in \mathcal{G}$, observation noise averages out and the sector gap $\Delta(A)$ can be recovered.
 \item Macroscopic consistency relies on the existence and invertibility of the physical link - mapping $T$ (e.g., energy-to-output laws). For evasion schemes that leave no macroscopic physical trace, $T$ is not invertible and macroscopic physical proxies fail.
 \item Rather than auditing every microscopic transaction in $\mathcal{F}$ via physical proxies, authorities can use coarse estimators on $\mathcal{G}$ to localize high-gap sectors and then deploy detailed audit within those sectors.
\end{itemize}

Next, we consider how the state can systematically increase microscopic resolution by fusing a sequence of distinct institutional measures/registries $\{\mu_1, \dots, \mu_n\}$.

\begin{Definition}[Fusion operator]
Let $\Phi_n$ be the cumulative institutional measure constructed from $\{\mu_1, \dots, \mu_n\}$. When a new institutional measure $\mu_{n+1}$ is introduced, let its Lebesgue decomposition with respect to $\Phi_n$ be
$$\mu_{n+1} = \mu^{ac}_{n+1} + \mu^{\perp}_{n+1},$$
where $\mu^{ac}_{n+1} \ll \Phi_n$ and $\mu^{\perp}_{n+1} \perp \Phi_n$ is supported on a measurable set $S^\perp \in \mathcal{F}$ with $\Phi_n(S^\perp) = 0$. The updated fusion measure $\Phi_{n+1}$ is defined for every $A \in \mathcal{F}$ by
\begin{equation}
 \Phi_{n+1}(A) = \Phi_n(A \setminus S^\perp) + \mu_{n+1}(A \cap S^\perp) + \int_A \max\left(0, \frac{d\mu_{n+1}^{ac}}{d\Phi_n} - 1\right) d\Phi_n,
 \label{Eq_FusionOperatorIntegral}
\end{equation}
which coincides with the Banach-lattice supremum, \cite{BanachLattice}, of the two measures
\begin{equation}
 \Phi_{n+1} = \Phi_n \vee \mu_{n+1}.
\end{equation}
\end{Definition}

Equation (\ref{Eq_FusionOperatorIntegral}) shows that a newly added registry $\mu_{n+1}$ expands state knowledge through two distinct channels: first, by its singular part $\mu_{n+1}^\perp$ on $S^\perp$ (uncovering previously unrecorded entities or transaction types where $\Phi_n = 0$), and second, by the excess density $\max\big(0, \frac{d\mu_{n+1}^{ac}}{d\Phi_n} - 1\big)$ on the already-monitored support (where $\mu_{n+1}$ observes a higher fraction of turnover than existing registries).
\begin{Theorem}[Asymptotic Reconstruction of True Measure]
\label{Th_AsymptoticReconstructionOfTrueMeasure}
Let $(\Omega, \mathcal{F}, \mu^*)$ be a finite measure space ($\mu^*(\Omega) < \infty$) representing the true economy, and let $\{\mu_k\}_{k=1}^\infty$ be a sequence of institutional measures satisfying $0 \le \mu_k \le \mu^*$ with Radon-Nikodym derivatives $g_k = \frac{d\mu_k}{d\mu^*}$ with values on $[0,1]$. Define the cumulative fusion operator for the first $n$ registries by
\begin{equation}
 \Phi_n = \bigvee_{k=1}^n \mu_k.
\end{equation}
Then $\{\Phi_n\}_{n=1}^\infty$ is a non-decreasing sequence of measures satisfying $\Phi_n \ll \mu^*$ with density $h_n(x) = \frac{d\Phi_n}{d\mu^*}(x) = \max_{1 \le k \le n} g_k(x) \in [0,1]$. Moreover, the following conditions hold:
\begin{enumerate}
 \item The sequence $\{\Phi_n\}$ converges to the true measure $\mu^*$ in total variation norm \cite{Folland, BanachLattice},
 \begin{equation}
  \lim_{n \to \infty} \Vert \mu^* - \Phi_n \Vert_{TV} = 0,
 \end{equation}
 so that the aggregate shadow economy vanishes ($\lim_{n \to \infty} \Delta_n = \lim_{n \to \infty} (\mu^*(\Omega) - \Phi_n(\Omega)) = 0$), if and only if the following condition holds:
 \begin{equation}
  \sup_{k \in \mathbb{N}} \frac{d\mu_k}{d\mu^*}(x) = 1 \quad \text{for } \mu^*\text{-almost all } x \in \Omega.
  \label{Eq_MaximalLocalCalibration}
 \end{equation}
 In particular, (\ref{Eq_MaximalLocalCalibration}) implies the following - defining the null supports $N_k = \{x \in \Omega \mid g_k(x) = 0\}$, we have
 \begin{equation}
  \mu^*\left(\bigcap_{k=1}^\infty N_k\right) = 0.
 \end{equation}
 \item Introducing a new institutional measure $\mu_{n+1}$ strictly reduces the unmeasured shadow-economy gap ($\Vert \mu^* - \Phi_{n+1}\Vert_{TV} < \Vert \mu^* - \Phi_n\Vert_{TV}$) if and only if
 \begin{equation}
  \mu^*\left(\{x \in \Omega \mid g_{n+1}(x) > h_n(x)\}\right) > 0,
 \end{equation}
 which holds if and only if, in the Lebesgue decomposition $\mu_{n+1} = \mu_{n+1}^{ac} + \mu_{n+1}^\perp$ with respect to $\Phi_n$, either $\mu_{n+1}^\perp \not\equiv 0$ (penetration of previous null/shadow economy sets) or $\Phi_n\left(\left\{x \in \Omega \mid \frac{d\mu_{n+1}^{ac}}{d\Phi_n}(x) > 1\right\}\right) > 0$ (improved visibility on already-monitored transactions).
\end{enumerate}
\end{Theorem}
\begin{Proof}[Proof Sketch]
The Radon-Nikodym derivative of $\Phi_n$ is $h_n(x) = \max_{1 \le k \le n} g_k(x)$. Since $1 - h_n(x)$ is a monotonically non-increasing sequence bounded by the integrable function $1$, convergence in total variation norm follows directly from applying the Lebesgue Dominated Convergence Theorem \cite{Folland}.
\end{Proof}

Theorem \ref{Th_AsymptoticReconstructionOfTrueMeasure} provides a blueprint for designing central fiscal databases: rather than building redundant registries that satisfy $\mu_{n+1} \le \Phi_n$, the Ministry of Finance should prioritize registries that penetrate existing null sets ($\mu_{n+1}^\perp \not\equiv 0$) or improve existing registers by capturing full transaction values where existing registries see only partial values. This contrasts with ecological Capture-Recapture (Petersen) methods \cite{CaputreRapture}, which rely on statistical independence and overlap between samples, whereas our fusion operator exploits mutual singularity and pointwise dominance, analogous to multi-sensor data fusion \cite{DataFusion}.

\subsection{Space of institutional measures}

The space of all finite signed measures on $(\Omega, \mathcal{F})$ forms a Banach space $\mathcal{M}(\Omega)$ equipped with the total variation norm $\Vert \cdot \Vert_{TV}$, see \cite{Folland}. The subspace of measures absolutely continuous with respect to $\mu^*$ is isometrically and order-isomorphic (via the Radon-Nikodym derivative $g_k = \frac{d\mu_k}{d\mu^*}$) to the Banach space $L^1(\Omega, \mathcal{F}, \mu^*)$. It leads to Banach lattice \cite{BanachLattice} when equipped with the pointwise lattice supremum ($\vee$) and infimum ($\wedge$) operations
\begin{eqnarray}
 (g_1 \vee g_2)(x) = \max(g_1(x), g_2(x)),\\
 (g_1 \wedge g_2)(x) = \min(g_1(x), g_2(x)).
\end{eqnarray}

For a countable collection of institutional measures $\mathcal{B} = \{\mu_k\}_{k=1}^\infty \subset L^1(\Omega, \mathcal{F}, \mu^*)$, we can analyze its completeness from two distinct functional-analytic perspectives: linear combinations (Banach space structure), and lattice suprema (Banach lattice structure).

In the linear (Data Fusion) approach, we approximate $\mu^*$ using conic linear combinations $\hat{\mu}_n = \sum_{k=1}^n w_k^{(n)} \mu_k$, $w_k^{(n)} \ge 0$. Distinguishing between the absence of common null sets (order-band completeness) and full linear density in $L^1(\mu^*)$ yields the following characterization.

\begin{Theorem}[Hahn-Banach and Topological Completeness for Shadow Economy]
\label{Th_Hahn-BanachTheoremForShadowEconomy}
Let $\mathcal{B} = \{\mu_k\}_{k=1}^\infty \subset L^1(\Omega, \mathcal{F}, \mu^*)$ be a countable collection of institutional measures with densities $g_k = \frac{d\mu_k}{d\mu^*} \in L^1(\mu^*)$. Then:
\begin{enumerate}
 \item Topological Completeness (Absence of Unobservable Domains): There is no non-zero shadow-economy domain $A \in \mathcal{F}$ ($\mu^*(A) > 0$) that is a simultaneous null set for all $\mu_k \in \mathcal{B}$ if and only if $\sup_{k \in \mathbb{N}} g_k(x) > 0$ $\mu^*$-a.e., which is equivalent to the condition that the closed order ideal (band) generated by $\mathcal{B}$ in the Banach lattice $L^1(\mu^*)$ equals the entire space $L^1(\Omega, \mathcal{F}, \mu^*)$.
 \item Linear Completeness: The true measure $\mu^*$ lies in the closed linear span $\overline{\text{span}(\mathcal{B})}$ in total variation norm, i.e.,
 \begin{equation}
  \lim_{n \to \infty} \left\Vert \mu^* - \sum_{k=1}^n w_k^{(n)} \mu_k \right\Vert_{TV} = 0,
 \end{equation}
 if and only if there exists no bounded measurable signed function $f \in L^\infty(\Omega, \mathcal{F}, \mu^*)$ that annihilates every institutional measure ($\int_\Omega f \, d\mu_k = 0$ for all $k \in \mathbb{N}$) while having non-zero integral of the true measure ($\int_\Omega f \, d\mu^* > 0$).
\end{enumerate}
\end{Theorem}
\begin{Proof}
The theorem is a classical result of Functional Analysis - for $L^1(\Omega, \mathcal{F}, \mu^*)$ case see \cite{Folland}, and for Banach lattice approach see \cite{BanachLattice}.
\end{Proof}

To estimate non-negative weights $w_1, \dots, w_n \ge 0$ in practice, one solves the $L^1$ regression problem
\begin{equation}
 \arg\min_{w_1, \dots, w_n \ge 0} \left\Vert \mu^* - \sum_{k=1}^n w_k \mu_k \right\Vert_{TV}.
\end{equation}
Note that we require $w_k \ge 0$ without forcing $\sum_{k=1}^n w_k = 1$, so that if individual registries systematically underreport turnover with $g_k < 1$, the weights can scale the linear combination up to match $\mu^*$.

In the second approach, we exploit the Banach lattice structure directly via the supremum fusion operator $\Phi_n = \bigvee_{k=1}^n \mu_k$. We have
\begin{Theorem}[Reconstruction of Full Measure by Fusion Operator]
\label{Th_ReconstructionByFusionOperator}
Let $\mathcal{B} = \{\mu_k\}_{k=1}^\infty$ be a countable family of institutional measures satisfying $0 \le \mu_k \le \mu^*$ with densities $g_k = \frac{d\mu_k}{d\mu^*}$ with values in $[0,1]$, and assume $\mu^*(\Omega) < \infty$. The sequence of fused measures $\Phi_n = \bigvee_{k=1}^n \mu_k$ converges to $\mu^*$ in total variation norm,
\begin{equation}
 \lim_{n \to \infty} \left\Vert \mu^* - \bigvee_{k=1}^n \mu_k \right\Vert_{TV} = 0,
\end{equation}
if and only if $\mathcal{B}$ satisfies the lattice faithfulness condition
\begin{equation}
 \sup_{\mu_k \in \mathcal{B}} \frac{d\mu_k}{d\mu^*}(x) = 1 \quad \mu^*\text{-a.e.}
\end{equation}
\end{Theorem}
\begin{Proof}
The proof is similar to the Theorem \ref{Th_AsymptoticReconstructionOfTrueMeasure} and bases on the properties of Banach lattice supremum.
\end{Proof}

Using the duality between $L^1$ and $L^\infty$, see \cite{Folland}, as measures measure the economy, the dual elements can be assigned to the economic strategy for economic transactions. We can interpret the previous theorem as follows.
\begin{Corollary}
A countable family of institutional measures $\mathcal{B} = \{\mu_1, \mu_2, \dots\}$ spans a dense subspace of $L^1(\Omega, \mathcal{F}, \mu^*)$,
\begin{equation}
 \overline{\text{span}(\mathcal{B})} = L^1(\Omega, \mathcal{F}, \mu^*),
\end{equation}
if and only if no non-zero signed evasion profile/strategy $f \in L^\infty(\Omega, \mathcal{F}, \mu^*)$ annihilates all $\mu_k \in \mathcal{B}$
\begin{equation}
 \left( \forall \mu_k \in \mathcal{B} : \int_\Omega f(x) \, d\mu_k(x) = 0 \right) \implies f(x) = 0 \quad \mu^*\text{-almost everywhere}.
\end{equation}
\end{Corollary}

Norm completeness of $L^1(\mu^*)$ guarantees that any Cauchy sequence of institutional estimators $\{\hat{\mu}_n\}_{n=1}^\infty$ ($\lim_{n,m \to \infty} \Vert \hat{\mu}_n - \hat{\mu}_m \Vert_{TV} = 0$) converges to a valid limit measure $\hat{\mu}_\infty \in L^1(\mu^*)$, while Dedekind completeness of the Banach lattice $L^1(\mu^*)$ ensures that the supremum $\bigvee_{\alpha} \mu_\alpha$ of any family of institutional measures bounded above by $\mu^*$ exists as a well-defined measure in $L^1(\mu^*)$ \cite{BanachLattice}.

\section{Hidden dimensions regime}
\label{Section_HiddenDimensionsRegime}

In this section we investigate a regime where the state possesses only a fixed, finite number of institutional registries $\{\mu_1, \dots, \mu_d\}$ ($d < \infty$) and cannot expand the registry base to cover all economic dimensions directly. Instead, institutional registries observe only marginal projections of economic activity (such as electricity consumption, registered employment, and freight volume), while other dimensions (such as off-the-books cash revenue or offshore settlements) remain unobserved.

We start our presentation from defining the setup for this regime. Let the total transaction space $\Omega$ (Borel space) be a product (Borel) space of $d$ observable coordinate spaces $\Omega_1, \dots, \Omega_d$ and a hidden coordinate (Borel) space $\Omega_{\text{hid}}$
\begin{eqnarray}
 \Omega_{\text{obs}} = \Omega_1 \times \dots \times \Omega_d, \\
 \Omega = \Omega_{\text{obs}} \times \Omega_{\text{hid}} = \Omega_1 \times \dots \times \Omega_d \times \Omega_{\text{hid}}.
\end{eqnarray}
Each transaction state is a vector $x = (x_{\text{obs}}, x_{\text{hid}}) \in \Omega$, where $x_{\text{obs}} = (x_1, \dots, x_d) \in \Omega_{\text{obs}}$. Normalization by total economic volume allows us to treat the true measure $\mu^* \in \mathcal{P}(\Omega)$ as a joint probability measure on $\Omega$. Defining the canonical coordinate projection operators $\pi_k: \Omega \to \Omega_k$ for $k = 1, \dots, d$, the $k$-th marginal institutional distribution is the pushforward measure
\begin{equation}
 \mu_k = \mu^* \circ \pi_k^{-1} \in \mathcal{P}(\Omega_k).
\end{equation}

First, consider the subproblem on the observable space $\Omega_{\text{obs}} = \mathbb{R}^d$ (or when $\Omega_{\text{hid}} = \emptyset$), where the state accurately observes the 1D marginals $\{\mu_1, \dots, \mu_d\}$ and seeks to reconstruct the joint distribution $\mu^*$.
\begin{Problem}[Marginal Reconstruction Problem]
Reconstruct the joint probability measure $\mu^*$ from a fixed finite set of marginal institutional measures $\{\mu_1, \dots, \mu_d\}$, $d < \infty$.
\end{Problem}

By Sklar's theorem \cite{Copula}, any joint distribution on $\mathbb{R}^d$ decomposes into its 1D marginals and a dependence structure (copula), which leads to the following theorem.
\begin{Theorem}[Sklar's Theorem for Shadow Economy (Theorem 2.10.9 in \cite{Copula})]
\label{Th_SklarTheoremShadowEconomy}
Every joint probability measure $\mu^*$ on $\mathbb{R}^d$ with cumulative distribution function $F_{\mu^*}$ and marginal cumulative distribution functions $F_{\mu_1}, \dots, F_{\mu_d}$ can be represented via a copula function $C: [0,1]^d \to [0,1]$ such that
\begin{equation}
 F_{\mu^*}(x_1, \dots, x_d) = C\big(F_{\mu_1}(x_1), \dots, F_{\mu_d}(x_d)\big).
 \label{Eq_SklarDistribution}
\end{equation}
If the marginals $F_{\mu_k}$ are continuous, the copula $C$ is unique.
\end{Theorem}

When $\mu^*$ and its marginals are absolutely continuous with respect to Lebesgue measure on $\mathbb{R}^d$ and $C$ admits a copula density $c(u_1, \dots, u_d) = \frac{\partial^d C(u_1, \dots, u_d)}{\partial u_1 \dots \partial u_d}$, the joint density $p^*$ factorizes into the product of marginal densities $p_k$ and the copula density $c$
\begin{equation}
 p^*(x_1, \dots, x_d) = c\big(F_{\mu_1}(x_1), \dots, F_{\mu_d}(x_d)\big) \prod_{k=1}^d p_k(x_k).
\end{equation}
This follows from differentiating (\ref{Eq_SklarDistribution}) under assumption that the copula and CDFs are sufficiently smooth, e.g., absolutely continuous.

Economically, the copula density $c$ encodes the \textit{technological coupling} between production inputs (e.g., labor, electricity, raw materials) and financial outputs. Tax evasion distorts the reported financial marginals relative to the joint distribution implied by the physical marginals and the technological copula.

To select a unique joint measure consistent with observed marginals and prior technological knowledge $\mu_{\text{tech}} \in \mathcal{P}(\Omega)$, we can select the procedure of minimizing the Kullback-Leibler (KL) divergence ($I$-projection of \cite{KLEntropy}).
\begin{Theorem}[Reconstruction via KL-divergence (\cite{KLEntropy})]
\label{Th_ReconstructionKL}
Let $\mu_{\text{tech}} \in \mathcal{P}(\Omega)$ be a reference technological prior measure, and let $\Gamma(\mu_1, \dots, \mu_d) = \{\nu \in \mathcal{P}(\Omega) \mid \nu \circ \pi_k^{-1} = \mu_k, \; k = 1, \dots, d\}$ be the Fr\'echet class \cite{MultivariateModels} of joint probability measures satisfying the marginal constraints. Assume there exists at least one feasible measure $\nu_0 \in \Gamma(\mu_1, \dots, \mu_d)$ such that $\text{KL}(\nu_0 \parallel \mu_{\text{tech}}) < \infty$. Then the optimization problem
\begin{equation}
 \hat{\mu}^* = \arg\min_{\nu \in \Gamma(\mu_1, \dots, \mu_d)} \text{KL}(\nu \parallel \mu_{\text{tech}}),
 \label{Eq_OptimizationProblemCopula}
\end{equation}
where
\begin{equation}
 \text{KL}(\nu \parallel \mu_{\text{tech}}) = \int_\Omega \log\left(\frac{d\nu}{d\mu_{\text{tech}}}\right) d\nu \quad \text{for } \nu \ll \mu_{\text{tech}},
\end{equation}
has a unique minimizer $\hat{\mu}^* \in \Gamma(\mu_1, \dots, \mu_d)$.

Moreover, when $\Omega$ is a finite discrete product space or when the densities are strictly positive and bounded away from zero and infinity on a rectangular support (so that the marginal log-potentials are integrable, $\lambda_k \in L^1(\mu_k)$ \cite{KLEntropy}), its Radon-Nikodym derivative with respect to $\mu_{\text{tech}}$ takes the exact multiplicative form
\begin{equation}
 d\hat{\mu}^*(x) = \left( \prod_{k=1}^d a_k(x_k) \right) d\mu_{\text{tech}}(x),
 \label{Eq_MultiAspectProductForm}
\end{equation}
where the marginal potentials $a_k(x_k):= \exp(\lambda_k(x_k) - 1/d) > 0$ are chosen so that $\hat{\mu}^* \circ \pi_k^{-1} = \mu_k$ for $k = 1, \dots, d$.
\end{Theorem}
\begin{Proof}[Proof Sketch]
The existence and uniqueness of the $I$-projection follow from the strict convexity of the Kullback-Leibler divergence on closed convex Fr\'echet constraint sets \cite{KLEntropy}. The multiplicative exponential form in equation (\ref{Eq_MultiAspectProductForm}) is a standard consequence of Csisz\'{a}r projection theorem via Lagrange multipliers for linear marginal constraints \cite{KLEntropy, MultivariateModels}.
\end{Proof}

We now ask when $\Omega = \Omega_{\text{obs}} \times \Omega_{\text{hid}}$, i.e., when $\Omega$ includes unobserved hidden dimensions $\Omega_{\text{hid}}$, under what exact mathematical conditions does the $I$-projection $\hat{\mu}^*$ recover the true joint measure $\mu^*$ everywhere on $\Omega$? Using measure disintegration and the chain rule for Kullback-Leibler divergence, we obtain a rigorous necessary and sufficient characterization.
\begin{Theorem}[Exact Reconstruction Condition (\cite{KLEntropy})]
\label{Th_ExactReconstruction}
Let $\Omega = \Omega_{\text{obs}} \times \Omega_{\text{hid}}$ with $\Omega_{\text{obs}} = \Omega_1 \times \dots \times \Omega_d$, and let $\Gamma = \{\nu \in \mathcal{P}(\Omega) \mid \nu \circ \pi_k^{-1} = \mu_k \text{ for } k = 1, \dots, d\}$ be the Fr\'echet class of measures matching the observable marginals $\mu_k = \mu^* \circ \pi_k^{-1}$. Assume $\text{KL}(\mu^* \parallel \mu_{\text{tech}}) < \infty$, and let
\begin{equation}
\hat{\mu}^* = \arg\min_{\nu \in \Gamma} \text{KL}(\nu \parallel \mu_{\text{tech}}).
\end{equation}

By disintegration of probability measures, write $d\nu(x_{\text{obs}}, x_{\text{hid}}) = \nu(dx_{\text{hid}} \mid x_{\text{obs}}) \, d\nu_{\text{obs}}(x_{\text{obs}})$, where $\nu_{\text{obs}}$ is the marginal of $\nu$ on $\Omega_{\text{obs}}$. Then the estimator recovers the true joint measure ($\hat{\mu}^* = \mu^*$) if and only if the following two conditions hold simultaneously:
\begin{enumerate}
 \item On the observable subspace $\Omega_{\text{obs}}$, the true observable marginal $\mu^*_{\text{obs}}$ is the $I$-projection of $\mu_{\text{tech, obs}}$ onto $\Gamma_{\text{obs}} = \{\eta \in \mathcal{P}(\Omega_{\text{obs}}) \mid \eta \circ \pi_k^{-1} = \mu_k\}$, which holds (under positive density support) if and only if the Radon-Nikodym derivative $\frac{d\mu^*_{\text{obs}}}{d\mu_{\text{tech, obs}}}$ factorizes into 1D marginal potentials
 \begin{equation}
  \frac{d\mu^*_{\text{obs}}}{d\mu_{\text{tech, obs}}}(x_1, \dots, x_d) = \prod_{k=1}^d a_k(x_k) \quad (\mu^*_{\text{obs}}\text{-a.e.}).
 \end{equation}
 That is, $\mu_{\text{tech, obs}}$ already captures the exact higher-order copula interactions among the observable dimensions $(X_1, \dots, X_d)$.
 \item For $\mu^*_{\text{obs}}$-almost every $x_{\text{obs}} \in \Omega_{\text{obs}}$, the conditional probability kernel of the true economy on $\Omega_{\text{hid}}$ coincides with the conditional technological prior
 \begin{equation}
  \mu^*(dx_{\text{hid}} \mid x_{\text{obs}}) = \mu_{\text{tech}}(dx_{\text{hid}} \mid x_{\text{obs}}).
 \end{equation}
 In particular, under existence of a measurable mapping $f: \Omega_{\text{obs}} \to \Omega_{\text{hid}}$ such that $X_{\text{hid}} = f(X_{\text{obs}})$ $\mu^*$-a.e. (corresponding on discrete partitions to zero conditional Shannon entropy $H(X_{\text{hid}} \mid X_{\text{obs}}) = 0$ \cite{ElementsOfInformaionTheory}, or for square-integrable random variables to zero conditional variance $\text{Var}(X_{\text{hid}} \mid X_{\text{obs}}) = 0$ a.e.) - exact recovery on $\Omega_{\text{hid}}$ holds if and only if the prior conditional kernel is concentrated on the deterministic graph of $f$, i.e., $\mu_{\text{tech}}(dx_{\text{hid}} \mid x_{\text{obs}}) = \delta_{f(x_{\text{obs}})}(dx_{\text{hid}})$.
\end{enumerate}
\end{Theorem}
\begin{Proof}
The proof is a repetition of the results of \cite{KLEntropy}, using in addition the existence of the mapping $f: \Omega_{\text{obs}} \to \Omega_{\text{hid}}$, \cite{ElementsOfInformaionTheory} that induces measure on $\Omega_{\text{hid}}$, and disintegration of the measure \cite{MeasureTheoryI_Bogachev}.
\end{Proof}

The condition 1 means that the government's baseline technological model already gets the joint relationships among the observable variables about total shadow economy right. In  condition 2 we assume that the government's baseline technological model already gets the conditional link from observable variables to hidden variables right. When condition 1 or 2 of Theorem \ref{Th_ExactReconstruction} fails - for instance, when the shadow economy adopts an innovative evasion copula that deviates from the 'business as usual' prior $\mu_{\text{tech}}$ - the reconstruction error $\text{KL}(\mu^* \parallel \hat{\mu}^*)$ is governed exactly by Csisz\'ar Information Pythagorean Theorem \cite{KLEntropy}.
\begin{Theorem}[The Information Pythagorean Theorem (Theorem 2.2 in \cite{KLEntropy})]
\label{Th_InformationPythagoreanTheorem}
Let $\mathcal{P}(\Omega)$ be the space of probability measures on $\Omega$, let $\mu_{\text{tech}} \in \mathcal{P}(\Omega)$ be a reference prior measure, and let $\Gamma \subset \mathcal{P}(\Omega)$ be a closed convex set of probability measures such that $\hat{\mu}^* = \arg\min_{\nu \in \Gamma} \text{KL}(\nu \parallel \mu_{\text{tech}})$ exists with $\text{KL}(\hat{\mu}^* \parallel \mu_{\text{tech}}) < \infty$. Then for any true measure $\mu^* \in \Gamma$ with $\text{KL}(\mu^* \parallel \mu_{\text{tech}}) < \infty$, the following inequality holds
\begin{equation}
 \text{KL}(\mu^* \parallel \mu_{\text{tech}}) \ge \text{KL}(\mu^* \parallel \hat{\mu}^*) + \text{KL}(\hat{\mu}^* \parallel \mu_{\text{tech}}).
 \label{Eq_InformationtriangleInequality}
\end{equation}
Furthermore, if $\Gamma = \Gamma(\mu_1, \dots, \mu_d)$ is defined by linear marginal constraints (with $\mu^* \ll \hat{\mu}^*$), the relationship holds with strict equality (see equation (3.1) in \cite{KLEntropy})
\begin{equation}
 \text{KL}(\mu^* \parallel \mu_{\text{tech}}) = \text{KL}(\mu^* \parallel \hat{\mu}^*) + \text{KL}(\hat{\mu}^* \parallel \mu_{\text{tech}}).
\end{equation}
The three terms can be given economic interpretation:
\begin{itemize}
 \item $\text{KL}(\mu^* \parallel \mu_{\text{tech}})$ - total information divergence between the true economy $\mu^*$ and the government's baseline technological prior $\mu_{\text{tech}}$.
 \item $\text{KL}(\hat{\mu}^* \parallel \mu_{\text{tech}})$ - the divergence accounted for by updating $\mu_{\text{tech}}$ to satisfy the observed institutional marginals $\mu_1, \dots, \mu_d$. Alternatively, this represents the information captured by the estimator under the assumed technological baseline.
 \item $\text{KL}(\mu^* \parallel \hat{\mu}^*)$ - the residual discrepancy between the true economy $\mu^*$ and the $I$-projection $\hat{\mu}^*$, caused by unmeasured cross-dimensional copula shifts or hidden-dimension deviations from $\mu_{\text{tech}}$.
\end{itemize}
\end{Theorem}

\begin{Corollary}
Since $I$-projection preserves the higher-order copula interactions and hidden-dimension conditional laws of the prior $\mu_{\text{tech}}$, any newly invented tax evasion scheme that alters the copula $C$ without changing the 1D marginals $\mu_k$ is invisible to the marginal constraints. Its entire divergence is absorbed into the unrecoverable error $\text{KL}(\mu^* \parallel \hat{\mu}^*)$. Furthermore, if $\mu^*$ assigns positive probability to a region where $\mu_{\text{tech}}$ vanishes ($\mu^* \not\ll \mu_{\text{tech}}$), then $\text{KL}(\mu^* \parallel \mu_{\text{tech}}) = +\infty$ and the prior must be recalibrated via ground-truth audits.
\end{Corollary}

\section{Compressed Sensing regime}
\label{Section_CS}

The regimes in Sections \ref{Section_LowResolutionRegime} and \ref{Section_HiddenDimensionsRegime} require either a sufficiently rich collection of registries (base measures) $\mathcal{B}$ or rigid technological coupling to physical inputs. In this section, we address the opposite regime. We consider an economy of $N$ entities where auditing every entity is too expensive, but high-volume tax evasion is concentrated among a sparse subset of $K \ll N$ actors (e.g., organized VAT carousel networks or concentrated illicit manufacturing).

We apply Compressed Sensing (CS) theory \cite{CSPaper1, CSPaper2, DataDrivenScience}, which has previously been used to detect electricity theft in smart grids \cite{CSInElectricityTheft2, CSInElectricityTheft} and anomalies in communication networks \cite{CS_InternetTraffic, CS_NetworksAnomaly}.

CS works in discrete setup. Therefore we discretize the transaction space $\Omega$ into a measurable partition of $N$, 'atoms' $\{A_j\}_{j=1}^N$, where $1 \le N < \infty$. Atoms reconstruct whole transaction space $\Omega = \bigsqcup_{j=1}^N A_j$, and are disjoint $A_i \cap A_j = \emptyset$ for $i \ne j$. Each atom $A_j$ can represents an individual firm, tax identification number, or local economic node.

We distinguish two measures on $(\Omega, \mathcal{F})$:
\begin{itemize}
 \item $\mu_{\text{leg}}$ (Legal Declared Measure) - the administrative measure induced by official tax filings and legal registries. By definition, $\mu_{\text{leg}}$ vanishes on completely unregistered shadow activities.
 \item $\mu_{\text{tech}}$ (Technological/Network Sensor Measure) - a physical or network-level sensing mechanism (such as aggregate smart-grid transformer loads, freight clearing nodes, or interbank liquidity settlements) that captures the true economic flow $\mu^*$ across linear combinations of atoms.
\end{itemize}

We define the true shadow-economy gap vector $s^* \in \mathbb{R}^N$ on the atoms $\{A_j\}_{j=1}^N$ by
\begin{equation}
 s^* = \begin{bmatrix} \mu^*(A_1) - \mu_{\text{leg}}(A_1) \\ \mu^*(A_2) - \mu_{\text{leg}}(A_2) \\ \vdots \\ \mu^*(A_N) - \mu_{\text{leg}}(A_N) \end{bmatrix},
 \label{Eq.SparseShadowEconomyVector}
\end{equation}
where $s^*_j = \mu_{\text{hid}}(A_j)$ for the signed or positive hidden measure $\mu_{\text{hid}} := \mu^* - \mu_{\text{leg}}$. Taking the Lebesgue decomposition of $\mu^*$ with respect to $\mu_{\text{leg}}$,
\begin{equation}
 \mu^* = r \, \mu_{\text{leg}} + \mu_{\text{sing}},
\end{equation}
where $r = \frac{d\mu^{*,ac}}{d\mu_{\text{leg}}} \ge 1$ is the Radon-Nikodym derivative of the absolutely continuous part of $\mu^*$ with respect to $\mu_{\text{leg}}$ and $\mu_{\text{sing}} \perp \mu_{\text{leg}}$ is supported on the combined shadow-economy null set $S_c$ (Definition \ref{Def_ShadowEconomyCombined}). Then each component $s^*_j$ decomposes into two evasion mechanisms
\begin{equation}
 s^*_j = \int_{A_j} (r(x) - 1) \, d\mu_{\text{leg}}(x) + \mu_{\text{sing}}(A_j).
\end{equation}
Here $\int_{A_j} (r(x) - 1) \, d\mu_{\text{leg}}(x)$ captures value underreporting on registered activities (information asymmetry), while $\mu_{\text{sing}}(A_j)$ captures completely off-the-books transactions.

Suppose the tax authority performs $M \ll N$ aggregated linear measurements $y \in \mathbb{R}^M$ using piecewise-constant sensing filters
\begin{equation}
 \phi_i(x) = \sum_{j=1}^N \Phi_{i,j} \chi_{A_j}(x), \quad i = 1, \dots, M.
\end{equation}
Then we can form Compressed Sensing equation
\begin{equation}
 y_i = \int_\Omega \phi_i(x) \, d\mu_{\text{hid}}(x) = \sum_{j=1}^N \Phi_{i,j} \mu_{\text{hid}}(A_j) = \sum_{j=1}^N \Phi_{i,j} s^*_j,
 \label{Eq_CSRecoveryEquation}
\end{equation}
or in matrix form
\begin{equation}
 y = \Phi s^*,
\end{equation}
where $\Phi \in \mathbb{R}^{M \times N}$ is the measurement (sensing) matrix. The equation connects the measured discrepancy $y$ with true shadow economy $s^*$. We are now ready for applying CS theory.
\begin{Theorem}[Exact Recovery of Sparse Shadow Economy (Theorem 1.2 in  \cite{CandesTaoLemma})]
\label{Th_CSRecovery}
Let $s^* \in \mathbb{R}^N$ be the unknown shadow-economy gap vector observed through $M \ll N$ aggregated checks $y = \Phi s^*$, where $\Phi \in \mathbb{R}^{M \times N}$. Assume:
\begin{enumerate}
 \item $K$-Sparsity in the Innovation Domain: The vector $s^*$ has at most $K$ non-zero entries ($K < M \ll N$):
 \begin{equation}
  \Vert s^* \Vert_0 = |\{j \in \{1, \dots, N\} \mid s^*_j \ne 0\}| \le K,
 \end{equation}
 where $\text{supp}(s^*) = \{j \mid s^*_j \ne 0\}$.
 \item Restricted Isometry Property (RIP) of Order $2K$: The sensing matrix $\Phi$ satisfies the RIP of order $2K$, meaning that for every $2K$-sparse vector $v \in \mathbb{R}^N$,
 \begin{equation}
  (1 - \delta_{2K}) \Vert v \Vert_2^2 \le \Vert \Phi v \Vert_2^2 \le (1 + \delta_{2K}) \Vert v \Vert_2^2,
 \end{equation}
 with restricted isometry constant strictly satisfying
 \begin{equation}
  \delta_{2K} < \sqrt{2} - 1.
 \end{equation}
\end{enumerate}
Then the convex $\ell_1$-minimization (Basis Pursuit) problem \cite{DataDrivenScience}
\begin{equation}
 \hat{s} = \arg\min_{s \in \mathbb{R}^N} \Vert s \Vert_1 \quad \text{subject to} \quad \Phi s = y
\end{equation}
has a unique solution, and recovers the true shadow economy vector exactly
\begin{equation}
 \hat{s} = s^*.
\end{equation}
\end{Theorem}
\begin{Proof}[Proof Sketch]
The proof adapts the restricted isometry property (RIP) and null-space geometry established by Cand\`es \cite{CandesTaoLemma, CS_RIP}. The error vector $h = \hat{s} - s^*$ satisfies the $\ell_1$ cone condition $\Vert h_{T_0^c} \Vert_1 \le \Vert h_{T_0} \Vert_1$, which geometrically traps perturbations within the kernel of $\Phi$. Applying the RIP condition $\delta_{2K} < \sqrt{2} - 1$ forces the projection norm $\Vert h_{T_0} \Vert_2$ to vanish, yielding exact recovery $\hat{s} = s^*$ \cite{CandesTaoLemma}.
\end{Proof}

Theorem \ref{Th_CSRecovery} yields two direct operational principles for organizing fiscal audits:
\begin{enumerate}
 \item When evasion is concentrated among $K \ll N$ entities, the state does not need $N$ individual audits; $M = \mathcal{O}(K \log(N/K))$ random incoherent aggregate checks suffice with high probability to satisfy the RIP of order $2K$ and reconstruct $s^*$ exactly \cite{CS_RIP, DataDrivenScience}.
 \item To satisfy the RIP, the sensing matrix $\Phi$ must be incoherent with the canonical entity basis, see detailed discussion of this property in \cite{DataDrivenScience}. Traditional siloed audits that inspect one industry or tax office at a time produce block-diagonal sensing matrices with poor restricted isometry constants. On the other hand, randomized cross-sectoral pooling is essential to detect sparse, high-volume fraud networks.
\end{enumerate}

We also emphasize that Theorem 1.3 in \cite{CandesTaoLemma} describes case where measurements are given by $y=\Phi s + z$, where the noise $z$ is bounded $\Vert z \Vert_{\ell_{2}}\leq \epsilon$. This allows to do reconstruction from noisy measurements.

We now provide examples where above methodology can be applied.
\begin{Example}[Currency Flow and Carousel Fraud Model]
\label{Ex_CS_CarouseleFraudModel}
Consider a network of $N$ firms where $K \ll N$ shell companies generate fictitious VAT invoices ($s^*$). Rather than auditing every invoice across all $N$ firms, $y \in \mathbb{R}^M$ records net VAT clearing imbalances aggregated across $M$ randomized multi-postal-code or multi-bank settlement pools defined by $\Phi$. Then CS methods allow to reconstruct true shadow economy $s^*$ since the shadow economy is sparse ($K \ll N$).
\end{Example}
\begin{Example}[Illegal Manufacturing via Smart-Grid Electricity Sensors]
\label{Ex_CS_ElectricityConsumption}
Let $N$ be the number of industrial plants in a region, and let $s^* \in \mathbb{R}^N$ denote unrecorded overproduction. The measurement vector $y \in \mathbb{R}^M$ records excess electricity draw (beyond declared production baselines) at $M$ substation feeders, where $\Phi_{i,j}$ converts the power draw of plant $j$ connected to feeder $i$ into monetary output equivalents via Leontief coefficients. When only $K \ll N$ plants engage in off-the-books production, $\ell_1$ minimization can pinpoint the exact plants from the feeder aggregates $y$ by looking on nonzero components of CS reconstructed $s^*$ vector.
\end{Example}
\begin{Example}[Informal Labor via Regional Cash Withdrawal Anomalies]
\label{Ex_CS_MoneyWithdrawn}
Let $N$ represent large employers in a metropolitan area and $s^*_j$ the total undeclared cash-in-hand wages paid by employer $j$. The vector $y \in \mathbb{R}^M$ measures payday cash withdrawal anomalies across $M$ ATM clusters, and $\Phi_{i,j}$ models the commuting probability from employer $j$ to ATM cluster $i$ derived from an urban mobility models, e.g., \cite{SmartCitiesTransportation}. If informal wage payouts are sparse ($K \ll N$), Basis Pursuit identifies the non-compliant employers.
\end{Example}

We want also comment on the inverse problem. In Compressed Sensing, one can also optimize the sensor placement matrix $\Phi$ using well-developed theory of Sparse Sensor Placement \cite{SparseSensorsPlacement1, SparseSensorsPlacement2, DataDrivenScience}. In the presence of observation noise $y = \Phi s^* + \epsilon_y$, reconstruction stability depends on the singular value spectrum of the $K$-column submatrices of $\Phi$: the ratio of the maximal to minimal singular values of $\Phi_T$ ($|T| \le 2K$), controlled by $\sqrt{(1+\delta_{2K})/(1-\delta_{2K})}$, should be as close to $1$ as possible. Calibrating an unknown physical transfer matrix $\Phi \in \mathbb{R}^{M \times N}$ empirically from historical audit pairs $(Y, S)$ via $Y = \Phi S$ requires $L \ge N$ linearly independent historical audit realizations $S = [s^{(1)}, \dots, s^{(L)}]$, and must account for behavioral shifts induced by the audits (noted in Section \ref{Section_IntuitivePicture}) that may alter $\Phi$ over time.

\section{Synthesis and Recommendations}

The three mathematical regimes developed in Sections \ref{Section_LowResolutionRegime}, \ref{Section_HiddenDimensionsRegime} and \ref{Section_CS} address complementary structural manifestations of the shadow economy. We started in Section \ref{Section_LowResolutionRegime} with the simplest case, in which the shadow economy was generated by underreporting and ineffective institutional measures. The solution is to add new 'orthogonal' institutional measures and use the fusion operator (lattice supremum of measures) to enhance sensitivity and approach the total measure. Then we expanded the model, in Section \ref{Section_HiddenDimensionsRegime}, to take into account the partition of measurements into the economic domain, and to the case where there is a part of the economy invisible to all institutional measures, however, visible to indirect measures based on physical activity (like energy consumption). Under certain assumptions about the technological processes involved, we can recover part of the invisible economy and combine it with estimates of the visible part based on the entropy maximization procedure. Finally, we consider the shadow economy generated by rare, high-volume actors in Section \ref{Section_CS}. The application of Compressed Sensing allowed us to be certain that, from selective, sparse, random measurements, we can recover information about these few agents.

These three cases represent the most typical situations and can be easily analyzed using standard theorems from measure theory, multivariate analysis, and compressed sensing. The crucial issue in the application is fulfilling the assumptions for the applicability of the underlying mathematical theory, which we formulated in economic terms.

We now provide recommendations for decision-makers that result from the above considered regimes. The important issue is to have a consistent picture of the economic transactions $\Omega$ from already implemented institutional measures. Therefore, tax authorities should link disparate state registries and apply the pointwise lattice supremum to establish an evidentiary lower bound on each taxpayer's true activity. Then, since the new institutional measure is a costly process, before funding a new national IT registry, policymakers should evaluate whether they gain truly new information about the economy from a pilot sample to verify that the new measure penetrates existing null sets or raises visibility above existing measures, avoiding ineffectiveness through costly duplication. If the institutional measure does not see hidden dimensions of the economy that can be traced by their physical fingerprints, then the introduction of a new institutional/technical measure must be based on a full understanding of the technical process that transforms the physical traces into economic transactions. Otherwise, the systematic error will be present. From a CS perspective, when targeting sparse, high-volume fraud networks, traditional sector-by-sector audits fail to satisfy the Restricted Isometry Property. Then, the best approach is to combine randomized cross-sectoral pooling of physical/network clearing flows with official tax declarations, allowing authorities to isolate and pinpoint major evaders via $\ell_1$ minimization at a fraction of the cost of traditional audits.

The final word is related to the Observer's paradox - the theory in this paper allows authorities to take a static snapshot of the full economy. However, it should be designed in a way that does not alarm participants of the shadow economy, since they can change their evasion strategies, leading to the rapid obsolescence of the snapshot.

\section{Conclusions}
We have established a formal measure-theoretic foundation for modeling and quantifying the shadow economy across three operational regimes: partial registry underreporting (Banach lattice theory), unrecorded economic dimensions (Copula theory and Maximal Entropy/$I$-projections), and sparse high-volume evasion networks (Compressed Sensing). For each regime, we proved necessary and sufficient conditions for the exact recovery of the true latent measure $\mu^*$ and formulated recommendations for decision-makers.

We highlight two primary theoretical limitations that define natural directions for future research. The first one concerns operating on a static snapshot. The present paper analyzes static measure spaces. Extending the framework to time-evolving shadow economies requires filtered probability spaces, stochastic processes, and martingale decompositions \cite{Measure_Schilling}. The other one is related to the observer's paradox. Since enforcement alters the underlying measure $\mu^*$ and its copula structure, a more advanced theory should account for endogenous changes in evasion strategies induced by measurement.

The presented theory can be used to introduce modern data-driven AI methods into GovTech applications, thereby introducing more explainability to the economy.

\section*{Acknowledgments}
RK work was partially funded by the EU Horizon Europe MSCA grant No 101300007.

During the preparation of this work, the authors used generative AI tools to assist with language polishing and structuring the presentation of proofs. After using these tools, the authors independently verified all mathematical derivations and take full responsibility for the content of the publication.

\appendix

\end{document}


{\LARGE\centering{\textbf{Supplement \\to\\'Shadow economy as a null set. Measure-theoretic approach to shadow economy'}}}

\vspace{0.4cm}
\begin{center}
Rados\l{}aw A. Kycia $^{a}$ \orcidlink{0000-0002-6390-4627},
Andrzej Buszko $^{b}$ \orcidlink{0000-0003-0600-4646}
\end{center}

\begin{center}
$^a$ Institute of Mathematical Economics, Collegium of Economic Analysis, Warsaw School of Economics, Warsaw, Poland \\
$^b$ Pomeranian University in S\l{}upsk, S\l{}upsk, Poland
\end{center}

\begin{abstract}
\noindent
In this supplement we provide proofs from the paper. They are based on adoption of standard results from measure theory, multivariate calculus and compressed sensing and therefore they are deferred to Supplement.
\end{abstract}

\noindent

\section{Low resolution regime}
\label{Section_LowResolutionRegime}
\begin{Theorem}[Optimal condition for institutional measure]
Suppose the operational cost $c: \Omega \times [0,1] \to \mathbb{R}_+$ of enforcing visibility $g \in [0,1]$ for transaction $x \in \Omega$ is a Carath\'eodory function (measurable in $x$ for each $g \in [0,1]$, and convex and continuously differentiable with respect to $g \in [0,1]$ for $\nu^*$-almost every $x \in \Omega$) satisfying $\sup_{g \in [0,1]} c(x,g) \in L^1(\Omega, \mathcal{F}, \nu^*)$. Given a tax rate $\tau \in [0,1]$ and $\nu^*$-integrable added-value function $f(x)$, a measurable visibility profile $g^*(x) \in [0,1]$ solves the net budget revenue maximization problem
\begin{equation}
  \max_{g: \Omega \to [0,1]} \int_{\Omega} \left( \tau f(x) g(x) - c(x, g(x)) \right) d\nu^*(x)
  \label{Eq.OptimalMeasure}
\end{equation}
if and only if for $\nu^*$-almost every $x \in \Omega$ we have
\begin{equation}
  \begin{cases}
    \tau f(x) \le \frac{\partial c}{\partial g}(x, 0) & \text{if } g^*(x) = 0, \\[4pt]
    \tau f(x) = \frac{\partial c}{\partial g}(x, g^*(x)) & \text{if } g^*(x) \in (0, 1), \\[4pt]
    \tau f(x) \ge \frac{\partial c}{\partial g}(x, 1) & \text{if } g^*(x) = 1.
  \end{cases}
  \label{Eq_KKT_Visibility}
\end{equation}
Here $\frac{\partial c}{\partial g}(x, g(x))$ represents the marginal enforcement cost of raising transaction visibility at $x$.
\end{Theorem}
\begin{Proof}
The proof is an application of Theorem 18.19 from \cite{InfiniteDimensionalAnalysis}.

Since $c(x,g)$ is a Carath\'eodory function with an integrable majorant and the integrand in (\ref{Eq.OptimalMeasure}) depends pointwise on $g(x) \in [0,1]$, it guarantees that maximizing the integral over measurable functions $g: \Omega \to [0,1]$ reduces (see also Theorem 14.60 of \cite{VariationalAnalysis_RockafellarWets}) to maximizing the pointwise objective function $I_x(g) = \tau f(x) g - c(x, g)$ over the compact interval $g \in [0,1]$ for $\nu^*$-almost every $x \in \Omega$.

Since $g \mapsto c(x, g)$ is convex and continuously differentiable on $[0,1]$, $I_x(g)$ is concave and continuously differentiable on $[0,1]$. By the Kuhn-Tucker optimality conditions on $[0,1]$, $g^* \in [0,1]$ maximizes $I_x(g)$ if and only if $I_x'(0) \le 0$ when $g^* = 0$, $I_x'(g^*) = 0$ when $g^* \in (0,1)$, and $I_x'(1) \ge 0$ when $g^* = 1$, which yields (\ref{Eq_KKT_Visibility}).
\end{Proof}
\begin{Corollary}
If $c(x,g)$ is convex with respect to $g \in [0,1]$, then for every transaction $x \in \Omega$ satisfying $\tau f(x) < \frac{\partial c}{\partial g}(x, 0)$, the unique optimal visibility in (\ref{Eq.OptimalMeasure}) is $g^*(x) = 0$.
\end{Corollary}
\begin{Proof}
For fixed $x$, define $I_x(g) = \tau f(x) g - c(x,g)$. Then $I_x'(g) = \tau f(x) - \frac{\partial c}{\partial g}(x,g)$. By convexity of $c(x, \cdot)$, the marginal cost $\frac{\partial c}{\partial g}(x,g)$ is non-decreasing in $g \in [0,1]$, so $\frac{\partial c}{\partial g}(x,0) \le \frac{\partial c}{\partial g}(x,g)$ for all $g \in [0,1]$. Since $\tau f(x) < \frac{\partial c}{\partial g}(x,0)$, we have $I_x'(g) < 0$ for all $g \in [0,1]$. Thus $I_x(g)$ is strictly decreasing on $[0,1]$ and attains its unique maximum at the boundary $g^*(x) = 0$.
\end{Proof}

\subsection{Estimation of total measure}

\begin{Theorem}[Convergence of Coarse Measures]
\label{Th_ConvergenceOfCoarseMeasures}
Let $(\Omega, \mathcal{F})$ be the measurable space of economic transactions, and let $\mathcal{G} \subsetneq \mathcal{F}$ be a coarse sub-$\sigma$-algebra generated by macroeconomic aggregates (sectors). Let $(\Theta, \Sigma, P)$ be an external probability space on which the true measure $\mu^*(\theta, \cdot)$ is drawn from a Bayesian prior distribution over finite positive measures on $(\Omega, \mathcal{F})$, and let $\mu \ll \mu^*$ be an institutional measure.

For each macroscopic sector $A \in \mathcal{G}$, suppose we observe a sequence of noisy macroscopic physical traces
\begin{equation}
 Z_i(A) = T(\mu^*(A)) + \epsilon_i(A), \quad i = 1, \dots, n,
\end{equation}
where $T: \mathbb{R}_+ \to \mathbb{R}$ is a continuous bijection with a continuous inverse $T^{-1}$, and $\{\epsilon_i(A)\}_{i=1}^\infty$ are pairwise uncorrelated random variables on $(\Theta, \Sigma, P)$ independent of $\mu^*$, with $E[\epsilon_i(A)] = 0$ and $\text{Var}(\epsilon_i(A)) = \sigma^2 \in \mathbb{R}_+$.

Define the macroscopic empirical estimator on $A \in \mathcal{G}$ by
\begin{equation}
 \hat{\mu}^*_n(A) := T^{-1}\left(\frac{1}{n}\sum_{i=1}^n Z_i(A)\right).
\end{equation}
Let
\begin{equation}
 \mathcal{H}_n = \sigma(\{Z_i(A) : A \in \mathcal{G}, \, 1 \le i \le n\}) \subset \Sigma
\end{equation}
denote the $\sigma$-algebra generated by all macroscopic observations up to sample size $n$, and let
\begin{equation}
 \mathcal{H}_\infty = \sigma(\{\mu^*(A) : A \in \mathcal{G}\}) \subset \Sigma.
\end{equation}
Then the following conditions hold:
\begin{enumerate}
 \item For any microeconomic subset $B \in \mathcal{F} \setminus \mathcal{G}$ such that $\mu^*(B) \in L^2(\Theta, \Sigma, P)$ has non-degenerate conditional prior variance given the macroscopic aggregates, i.e., $c := E[\text{Var}(\mu^*(B) \mid \mathcal{H}_\infty)] > 0$, any sequence of estimators $\tilde{\mu}_n(B)$ measurable with respect to the macroscopic observations $\mathcal{H}_n$ has mean squared error bounded away from zero, i.e.,
 \begin{equation}
  \inf_{n \in \mathbb{N}} E\left[ \left(\tilde{\mu}_n(B) - \mu^*(B)\right)^2 \right] \ge c > 0.
 \end{equation}
 \item Conditional on any realization of $\mu^*$, for every macroscopic aggregate $A \in \mathcal{G}$, the estimator $\hat{\mu}^*_n(A)$ converges in probability to the true sector value $\mu^*(A)$ as $n \to \infty$, namely,
 \begin{equation}
  \forall \epsilon > 0, \quad \lim_{n \to \infty} P\left( \left\vert \hat{\mu}^*_n(A) - \mu^*(A) \right\vert > \epsilon \;\middle|\; \mu^*(A) \right) = 0.
 \end{equation}
 Consequently, the macroeconomic shadow-economy estimator $\hat{\Delta}_n(A) = \hat{\mu}^*_n(A) - \mu(A)$ is consistent on $\mathcal{G}$, i.e.,
 \begin{equation}
  \text{plim}_{n \to \infty} \hat{\Delta}_n(A) = \Delta(A).
 \end{equation}
\end{enumerate}
\end{Theorem}
\begin{Proof}
We start from second statement. Fix a macroscopic sector $A \in \mathcal{G}$ and condition on the realization of $\mu^*(A)$. The empirical average of the physical trace observations is
\begin{equation}
\bar{Z}_n(A) = \frac{1}{n} \sum_{i=1}^n Z_i(A) = T(\mu^*(A)) + \frac{1}{n} \sum_{i=1}^n \epsilon_i(A).
\end{equation}
Since the noise variables $\{\epsilon_i(A)\}_{i=1}^\infty$ are pairwise uncorrelated with $E[\epsilon_i(A)] = 0$ and common finite variance $\sigma^2$, Chebyshev's inequality (Weak Law of Large Numbers \cite{Folland}) gives
\begin{equation}
 \text{plim}_{n \to \infty} \frac{1}{n}\sum_{i=1}^n \epsilon_i(A) = 0, \quad \text{so} \quad \bar{Z}_n(A) \xrightarrow{P} T(\mu^*(A)).
\end{equation}
Because $T^{-1}$ is continuous, the Continuous Mapping Theorem \cite{Folland} implies
$$\hat{\mu}^*_n(A) = T^{-1}(\bar{Z}_n(A)) \xrightarrow{P} T^{-1}(T(\mu^*(A))) = \mu^*(A),$$
and by Slutsky's theorem (respecting limit in the probability by continuous operations \cite{Folland}),
\begin{equation}
 \hat{\Delta}_n(A) = \hat{\mu}^*_n(A) - \mu(A) \xrightarrow{P} \mu^*(A) - \mu(A) = \Delta(A).
\end{equation}

We now prove the first statement. Select a microeconomic subset $B \in \mathcal{F} \setminus \mathcal{G}$ and let $\tilde{\mu}_n(B)$ be any $\mathcal{H}_n$-measurable estimator. Define the noise $\sigma$-algebra $\mathcal{E} = \sigma(\{\epsilon_i(A) : A \in \mathcal{G}, \, i \ge 1\}) \subset \Sigma$. Since each $Z_i(A)$ is a deterministic function of $\mu^*(A)$ and $\epsilon_i(A)$, we have $\mathcal{H}_n \subset \mathcal{H}_\infty \vee \mathcal{E}$ for every $n \in \mathbb{N}$.

Because the observation noise $\mathcal{E}$ is independent of $\mu^*$, the $L^2(P)$-orthogonal projection (Section 9 of \cite{ProbabilityWithMartingales} or Section 27 of \cite{Measure_Schilling}) of $\mu^*(B)$ onto $L^2(\Theta, \mathcal{H}_\infty \vee \mathcal{E}, P)$ is simply $\hat{\mu}_{\text{opt}}(B) = E[\mu^*(B) \mid \mathcal{H}_\infty \vee \mathcal{E}] = E[\mu^*(B) \mid \mathcal{H}_\infty]$. By the law of total variance and the $L^2$-optimality of conditional expectation, for every $n \in \mathbb{N}$
\begin{eqnarray}
 E\left[\left(\tilde{\mu}_n(B) - \mu^*(B)\right)^2\right] \ge E\left[\left(E[\mu^*(B) \mid \mathcal{H}_\infty] - \mu^*(B)\right)^2\right] = \notag \\
 =E\left[\text{Var}(\mu^*(B) \mid \mathcal{H}_\infty)\right] = c > 0.
\end{eqnarray}
Taking the infimum over $n \in \mathbb{N}$ yields $\inf_{n \in \mathbb{N}} E[(\tilde{\mu}_n(B) - \mu^*(B))^2] \ge c > 0$. Thus, collecting arbitrarily many macroscopic observations cannot resolve the microeconomic allocation of the shadow economy within $\mathcal{F} \setminus \mathcal{G}$.
\end{Proof}
\begin{Theorem}[Asymptotic Reconstruction of True Measure]
\label{Th_AsymptoticReconstructionOfTrueMeasure}
Let $(\Omega, \mathcal{F}, \mu^*)$ be a finite measure space ($\mu^*(\Omega) < \infty$) representing the true economy, and let $\{\mu_k\}_{k=1}^\infty$ be a sequence of institutional measures satisfying $0 \le \mu_k \le \mu^*$ with Radon-Nikodym derivatives $g_k = \frac{d\mu_k}{d\mu^*}$ with values on $[0,1]$. Define the cumulative fusion operator for the first $n$ registries by
\begin{equation}
 \Phi_n = \bigvee_{k=1}^n \mu_k.
\end{equation}
Then $\{\Phi_n\}_{n=1}^\infty$ is a non-decreasing sequence of measures satisfying $\Phi_n \ll \mu^*$ with density $h_n(x) = \frac{d\Phi_n}{d\mu^*}(x) = \max_{1 \le k \le n} g_k(x) \in [0,1]$. Moreover, the following conditions hold:
\begin{enumerate}
 \item The sequence $\{\Phi_n\}$ converges to the true measure $\mu^*$ in total variation norm \cite{Folland, BanachLattice},
 \begin{equation}
  \lim_{n \to \infty} \Vert \mu^* - \Phi_n \Vert_{TV} = 0,
 \end{equation}
 so that the aggregate shadow economy vanishes ($\lim_{n \to \infty} \Delta_n = \lim_{n \to \infty} (\mu^*(\Omega) - \Phi_n(\Omega)) = 0$), if and only if the following condition holds:
 \begin{equation}
  \sup_{k \in \mathbb{N}} \frac{d\mu_k}{d\mu^*}(x) = 1 \quad \text{for } \mu^*\text{-almost all } x \in \Omega.
  \label{Eq_MaximalLocalCalibration}
 \end{equation}
 In particular, (\ref{Eq_MaximalLocalCalibration}) implies the following - defining the null supports $N_k = \{x \in \Omega \mid g_k(x) = 0\}$, we have
 \begin{equation}
  \mu^*\left(\bigcap_{k=1}^\infty N_k\right) = 0.
 \end{equation}
 \item Introducing a new institutional measure $\mu_{n+1}$ strictly reduces the unmeasured shadow-economy gap ($\Vert \mu^* - \Phi_{n+1}\Vert_{TV} < \Vert \mu^* - \Phi_n\Vert_{TV}$) if and only if
 \begin{equation}
  \mu^*\left(\{x \in \Omega \mid g_{n+1}(x) > h_n(x)\}\right) > 0,
 \end{equation}
 which holds if and only if, in the Lebesgue decomposition $\mu_{n+1} = \mu_{n+1}^{ac} + \mu_{n+1}^\perp$ with respect to $\Phi_n$, either $\mu_{n+1}^\perp \not\equiv 0$ (penetration of previous null/shadow economy sets) or $\Phi_n\left(\left\{x \in \Omega \mid \frac{d\mu_{n+1}^{ac}}{d\Phi_n}(x) > 1\right\}\right) > 0$ (improved visibility on already-monitored transactions).
\end{enumerate}
\end{Theorem}
\begin{Proof}
Since $0 \le \mu_k \le \mu^*$ for all $k$, each $\Phi_n = \bigvee_{k=1}^n \mu_k$ satisfies $0 \le \Phi_n \le \Phi_{n+1} \le \mu^*$, so $\Phi_n \ll \mu^*$ with Radon-Nikodym derivative
\begin{equation}
 h_n(x) = \frac{d\Phi_n}{d\mu^*}(x) = \max_{1 \le k \le n} g_k(x) \in [0,1] \quad \text{for } \mu^*\text{-a.e. } x \in \Omega.
\end{equation}
Because $h_{n+1}(x) = \max(h_n(x), g_{n+1}(x))$, the sequence $\{h_n(x)\}_{n=1}^\infty$ is pointwise non-decreasing and bounded above by $1$, and therefore converges pointwise to $h_\infty(x) := \sup_{k \in \mathbb{N}} g_k(x) \in [0,1]$.

For measures $\Phi_n \le \mu^*$, the total variation distance is given by
\begin{equation}
 \Vert \mu^* - \Phi_n \Vert_{TV} = \int_\Omega \left\vert \frac{d\mu^*}{d\mu^*} - \frac{d\Phi_n}{d\mu^*} \right\vert d\mu^* = \int_\Omega (1 - h_n(x)) \, d\mu^*(x).
\end{equation}
Since $0 \le 1 - h_n(x) \le 1$ and the constant function $1$ is $\mu^*$-integrable (as $\mu^*(\Omega) < \infty$), the Lebesgue Dominated Convergence Theorem \cite{Folland} yields
\begin{equation}
 \lim_{n \to \infty} \Vert \mu^* - \Phi_n \Vert_{TV} = \lim_{n \to \infty} \int_\Omega (1 - h_n(x)) \, d\mu^*(x) = \int_\Omega (1 - h_\infty(x)) \, d\mu^*(x).
\end{equation}
Because $1 - h_\infty(x) \ge 0$ $\mu^*$-a.e., the integral on the right-hand side equals $0$ if and only if $h_\infty(x) = \sup_{k \in \mathbb{N}} g_k(x) = 1$ for $\mu^*$-a.e. $x \in \Omega$. Therefore, the common null support $\bigcap_{k=1}^\infty N_k = \{x \in \Omega \mid h_\infty(x) = 0\}$ is a subset of $\{x \in \Omega \mid h_\infty(x) < 1\}$ and therefore has $\mu^*$-measure zero.

To prove Statement 2, observe that at step $n+1$, the reduction in the total variation gap is
\begin{eqnarray}
 \Vert \mu^* - \Phi_n \Vert_{TV} - \Vert \mu^* - \Phi_{n+1} \Vert_{TV} = \int_\Omega (h_{n+1}(x) - h_n(x)) \, d\mu^*(x) = \notag \\
 =\int_\Omega \max(0, g_{n+1}(x) - h_n(x)) \, d\mu^*(x).
\end{eqnarray}
By the definition of Fusion Operator
\begin{equation*}
 \Phi_{n+1}(A) = \Phi_n(A \setminus S^\perp) + \mu_{n+1}(A \cap S^\perp) + \int_A \max\left(0, \frac{d\mu_{n+1}^{ac}}{d\Phi_n} - 1\right) d\Phi_n,
\end{equation*}
this integral equals $\mu_{n+1}^\perp(\Omega) + \int_\Omega \max\left(0, \frac{d\mu_{n+1}^{ac}}{d\Phi_n} - 1\right) d\Phi_n$. Since the integrand is non-negative, the gap strictly decreases if and only if $\mu^*(\{x \in \Omega \mid g_{n+1}(x) > h_n(x)\}) > 0$, which is equivalent to either $\mu_{n+1}^\perp(\Omega) > 0$ or $\Phi_n\left(\left\{x \in \Omega \mid \frac{d\mu_{n+1}^{ac}}{d\Phi_n}(x) > 1\right\}\right) > 0$.
\end{Proof}

\subsection{Space of institutional measures}

\begin{Theorem}[Hahn-Banach and Topological Completeness for Shadow Economy]
\label{Th_Hahn-BanachTheoremForShadowEconomy}
Let $\mathcal{B} = \{\mu_k\}_{k=1}^\infty \subset L^1(\Omega, \mathcal{F}, \mu^*)$ be a countable collection of institutional measures with densities $g_k = \frac{d\mu_k}{d\mu^*} \in L^1(\mu^*)$. Then:
\begin{enumerate}
 \item Topological Completeness (Absence of Unobservable Domains): There is no non-zero shadow-economy domain $A \in \mathcal{F}$ ($\mu^*(A) > 0$) that is a simultaneous null set for all $\mu_k \in \mathcal{B}$ if and only if $\sup_{k \in \mathbb{N}} g_k(x) > 0$ $\mu^*$-a.e., which is equivalent to the condition that the closed order ideal (band) generated by $\mathcal{B}$ in the Banach lattice $L^1(\mu^*)$ equals the entire space $L^1(\Omega, \mathcal{F}, \mu^*)$.
 \item Linear Completeness: The true measure $\mu^*$ lies in the closed linear span $\overline{\text{span}(\mathcal{B})}$ in total variation norm, i.e.,
 \begin{equation}
  \lim_{n \to \infty} \left\Vert \mu^* - \sum_{k=1}^n w_k^{(n)} \mu_k \right\Vert_{TV} = 0,
 \end{equation}
 if and only if there exists no bounded measurable signed function $f \in L^\infty(\Omega, \mathcal{F}, \mu^*)$ that annihilates every institutional measure ($\int_\Omega f \, d\mu_k = 0$ for all $k \in \mathbb{N}$) while having non-zero integral of the true measure ($\int_\Omega f \, d\mu^* > 0$).
\end{enumerate}
\end{Theorem}
\begin{Proof}
We start from the proof of first statement. A measurable set $A \in \mathcal{F}$ satisfies $\mu_k(A) = \int_A g_k \, d\mu^* = 0$ for all $k \in \mathbb{N}$ if and only if $g_k(x) = 0$ $\mu^*$-a.e. on $A$ for all $k$, i.e., $\sup_{k \in \mathbb{N}} g_k(x) = 0$ $\mu^*$-a.e. on $A$. Thus no such set with $\mu^*(A) > 0$ exists if and only if $\sup_{k \in \mathbb{N}} g_k(x) > 0$ $\mu^*$-a.e. By standard Banach lattice theory \cite{BanachLattice}, in $L^1(\mu^*)$ the disjoint complement $\mathcal{B}^d = \{f \in L^1(\mu^*) \mid |f| \wedge g_k = 0, \;\forall k\}$ is trivial ($\mathcal{B}^d = \{0\}$), and hence the band $\mathcal{B}^{dd}$ generated by $\mathcal{B}$ equals $L^1(\mu^*)$, if and only if $\sup_{k \in \mathbb{N}} g_k(x) > 0$ $\mu^*$-a.e.

For the second claim, since the dual space of $L^1(\Omega, \mathcal{F}, \mu^*)$ is isometrically isomorphic to $L^\infty(\Omega, \mathcal{F}, \mu^*)$, the Hahn-Banach separation theorem \cite{Folland} states that $\mu^* \notin \overline{\text{span}(\mathcal{B})}$ if and only if there exists a continuous linear functional $f \in L^\infty(\Omega, \mathcal{F}, \mu^*)$ such that $\int_\Omega f \, d\mu_k = \int_\Omega f g_k \, d\mu^* = 0$ for all $k \in \mathbb{N}$ and $\int_\Omega f \, d\mu^* > 0$.
\end{Proof}
\begin{Theorem}[Reconstruction of Full Measure by Fusion Operator]
\label{Th_ReconstructionByFusionOperator}
Let $\mathcal{B} = \{\mu_k\}_{k=1}^\infty$ be a countable family of institutional measures satisfying $0 \le \mu_k \le \mu^*$ with densities $g_k = \frac{d\mu_k}{d\mu^*}$ with values in $[0,1]$, and assume $\mu^*(\Omega) < \infty$. The sequence of fused measures $\Phi_n = \bigvee_{k=1}^n \mu_k$ converges to $\mu^*$ in total variation norm,
\begin{equation}
 \lim_{n \to \infty} \left\Vert \mu^* - \bigvee_{k=1}^n \mu_k \right\Vert_{TV} = 0,
\end{equation}
if and only if $\mathcal{B}$ satisfies the lattice faithfulness condition
\begin{equation}
 \sup_{\mu_k \in \mathcal{B}} \frac{d\mu_k}{d\mu^*}(x) = 1 \quad \mu^*\text{-a.e.}
\end{equation}
\end{Theorem}
\begin{Proof}
The Radon-Nikodym derivative of $\Phi_n = \bigvee_{k=1}^n \mu_k$ with respect to $\mu^*$ is $G_n(x) = \max_{1 \le k \le n} g_k(x) \in [0,1]$, and
\begin{equation}
 \left\Vert \mu^* - \Phi_n \right\Vert_{TV} = \int_{\Omega} (1 - G_n(x)) \, d\mu^*(x).
\end{equation}
Since $1 - G_n(x) \ge 0$ is monotonically non-increasing and bounded above by the $\mu^*$-integrable function $1$, the Lebesgue Dominated Convergence Theorem, \cite{Folland}, gives
\begin{equation}
 \lim_{n \to \infty} \left\Vert \mu^* - \Phi_n \right\Vert_{TV} = \int_{\Omega} \left(1 - \sup_{\mu_k \in \mathcal{B}} g_k(x)\right) d\mu^*(x).
\end{equation}
Because the integrand $1 - \sup_{\mu_k \in \mathcal{B}} g_k(x)$ is non-negative $\mu^*$-a.e., its integral vanishes if and only if $\sup_{\mu_k \in \mathcal{B}} g_k(x) = 1$ $\mu^*$-a.e.
\end{Proof}

\begin{Corollary}
A countable family of institutional measures $\mathcal{B} = \{\mu_1, \mu_2, \dots\}$ spans a dense subspace of $L^1(\Omega, \mathcal{F}, \mu^*)$,
\begin{equation}
 \overline{\text{span}(\mathcal{B})} = L^1(\Omega, \mathcal{F}, \mu^*),
\end{equation}
if and only if no non-zero signed evasion profile/strategy $f \in L^\infty(\Omega, \mathcal{F}, \mu^*)$ annihilates all $\mu_k \in \mathcal{B}$
\begin{equation}
 \left( \forall \mu_k \in \mathcal{B} : \int_\Omega f(x) \, d\mu_k(x) = 0 \right) \implies f(x) = 0 \quad \mu^*\text{-almost everywhere}.
\end{equation}
\end{Corollary}

\section{Hidden dimensions regime}
\label{Section_HiddenDimensionsRegime}

\begin{Theorem}[Sklar's Theorem for Shadow Economy (Theorem 2.10.9 in \cite{Copula})]
\label{Th_SklarTheoremShadowEconomy}
Every joint probability measure $\mu^*$ on $\mathbb{R}^d$ with cumulative distribution function $F_{\mu^*}$ and marginal cumulative distribution functions $F_{\mu_1}, \dots, F_{\mu_d}$ can be represented via a copula function $C: [0,1]^d \to [0,1]$ such that
\begin{equation}
 F_{\mu^*}(x_1, \dots, x_d) = C\big(F_{\mu_1}(x_1), \dots, F_{\mu_d}(x_d)\big).
 \label{Eq_SklarDistribution}
\end{equation}
If the marginals $F_{\mu_k}$ are continuous, the copula $C$ is unique.
\end{Theorem}
\begin{Proof}
 Theorem 2.10.9 in \cite{Copula}
\end{Proof}

\begin{Theorem}[Reconstruction via KL-divergence (\cite{KLEntropy})]
\label{Th_ReconstructionKL}
Let $\mu_{\text{tech}} \in \mathcal{P}(\Omega)$ be a reference technological prior measure, and let $\Gamma(\mu_1, \dots, \mu_d) = \{\nu \in \mathcal{P}(\Omega) \mid \nu \circ \pi_k^{-1} = \mu_k, \; k = 1, \dots, d\}$ be the Fr\'echet class \cite{MultivariateModels} of joint probability measures satisfying the marginal constraints. Assume there exists at least one feasible measure $\nu_0 \in \Gamma(\mu_1, \dots, \mu_d)$ such that $\text{KL}(\nu_0 \parallel \mu_{\text{tech}}) < \infty$. Then the optimization problem
\begin{equation}
 \hat{\mu}^* = \arg\min_{\nu \in \Gamma(\mu_1, \dots, \mu_d)} \text{KL}(\nu \parallel \mu_{\text{tech}}),
 \label{Eq_OptimizationProblemCopula}
\end{equation}
where
\begin{equation}
 \text{KL}(\nu \parallel \mu_{\text{tech}}) = \int_\Omega \log\left(\frac{d\nu}{d\mu_{\text{tech}}}\right) d\nu \quad \text{for } \nu \ll \mu_{\text{tech}},
\end{equation}
has a unique minimizer $\hat{\mu}^* \in \Gamma(\mu_1, \dots, \mu_d)$.

Moreover, when $\Omega$ is a finite discrete product space or when the densities are strictly positive and bounded away from zero and infinity on a rectangular support (so that the marginal log-potentials are integrable, $\lambda_k \in L^1(\mu_k)$ \cite{KLEntropy}), its Radon-Nikodym derivative with respect to $\mu_{\text{tech}}$ takes the exact multiplicative form
\begin{equation}
 d\hat{\mu}^*(x) = \left( \prod_{k=1}^d a_k(x_k) \right) d\mu_{\text{tech}}(x),
 \label{Eq_MultiAspectProductForm}
\end{equation}
where the marginal potentials $a_k(x_k):= \exp(\lambda_k(x_k) - 1/d) > 0$ are chosen so that $\hat{\mu}^* \circ \pi_k^{-1} = \mu_k$ for $k = 1, \dots, d$.
\end{Theorem}
\begin{Proof}
The Fr\'echet constraint set $\Gamma(\mu_1, \dots, \mu_d)$ defined by the continuous linear pushforward operators $\nu \mapsto \nu \circ \pi_k^{-1}$ is convex and closed in total variation norm. Since $\nu \mapsto \text{KL}(\nu \parallel \mu_{\text{tech}})$ is strictly convex on $\{\nu \in \mathcal{P}(\Omega) \mid \text{KL}(\nu \parallel \mu_{\text{tech}}) < \infty\}$ and has compact sublevel sets in the weak topology, the existence of $\nu_0 \in \Gamma(\mu_1, \dots, \mu_d)$ with $\text{KL}(\nu_0 \parallel \mu_{\text{tech}}) < \infty$ guarantees the existence and uniqueness of the global minimizer ($I$-projection) $\hat{\mu}^* \in \Gamma(\mu_1, \dots, \mu_d)$ \cite{KLEntropy, MultivariateModels}.

To derive the functional form (\ref{Eq_MultiAspectProductForm}) via Lagrange multipliers $\lambda_k: \Omega_k \to \mathbb{R}$, form the Lagrangian functional
\begin{eqnarray}
 \mathcal{L}(\nu, \lambda_1, \dots, \lambda_d) = \int_{\Omega} \log\left(\frac{d\nu}{d\mu_{\text{tech}}}(x)\right) d\nu(x) + \notag \\
 - \sum_{k=1}^d \left( \int_{\Omega} \lambda_k(x_k) \, d\nu(x) - \int_{\Omega_k} \lambda_k(x_k) \, d\mu_k(x_k) \right).
\end{eqnarray}
Taking the first variation with respect to $\delta\nu$ gives
\begin{equation}
 \delta \mathcal{L} = \int_{\Omega} \left[ \log\left(\frac{d\nu}{d\mu_{\text{tech}}}(x)\right) + 1 - \sum_{k=1}^d \lambda_k(x_k) \right] d(\delta \nu)(x) = 0.
\end{equation}
Setting the integrand in brackets to zero $\mu_{\text{tech}}$-a.e. yields
\begin{equation}
 \frac{d\hat{\mu}^*}{d\mu_{\text{tech}}}(x) = \exp\left( \sum_{k=1}^d \lambda_k(x_k) - 1 \right) = \prod_{k=1}^d a_k(x_k),
\end{equation}
where $a_k(x_k) := \exp(\lambda_k(x_k) - 1/d)$. Imposing the marginal constraints $\hat{\mu}^* \circ \pi_k^{-1} = \mu_k$ determines the potentials $a_k(x_k)$.
\end{Proof}
\begin{Theorem}[Exact Reconstruction Condition (\cite{KLEntropy})]
\label{Th_ExactReconstruction}
Let $\Omega = \Omega_{\text{obs}} \times \Omega_{\text{hid}}$ with $\Omega_{\text{obs}} = \Omega_1 \times \dots \times \Omega_d$, and let $\Gamma = \{\nu \in \mathcal{P}(\Omega) \mid \nu \circ \pi_k^{-1} = \mu_k \text{ for } k = 1, \dots, d\}$ be the Fr\'echet class of measures matching the observable marginals $\mu_k = \mu^* \circ \pi_k^{-1}$. Assume $\text{KL}(\mu^* \parallel \mu_{\text{tech}}) < \infty$, and let
\begin{equation}
\hat{\mu}^* = \arg\min_{\nu \in \Gamma} \text{KL}(\nu \parallel \mu_{\text{tech}}).
\end{equation}

By disintegration of probability measures, write $d\nu(x_{\text{obs}}, x_{\text{hid}}) = \nu(dx_{\text{hid}} \mid x_{\text{obs}}) \, d\nu_{\text{obs}}(x_{\text{obs}})$, where $\nu_{\text{obs}}$ is the marginal of $\nu$ on $\Omega_{\text{obs}}$. Then the estimator recovers the true joint measure ($\hat{\mu}^* = \mu^*$) if and only if the following two conditions hold simultaneously:
\begin{enumerate}
 \item On the observable subspace $\Omega_{\text{obs}}$, the true observable marginal $\mu^*_{\text{obs}}$ is the $I$-projection of $\mu_{\text{tech, obs}}$ onto $\Gamma_{\text{obs}} = \{\eta \in \mathcal{P}(\Omega_{\text{obs}}) \mid \eta \circ \pi_k^{-1} = \mu_k\}$, which holds (under positive density support) if and only if the Radon-Nikodym derivative $\frac{d\mu^*_{\text{obs}}}{d\mu_{\text{tech, obs}}}$ factorizes into 1D marginal potentials
 \begin{equation}
  \frac{d\mu^*_{\text{obs}}}{d\mu_{\text{tech, obs}}}(x_1, \dots, x_d) = \prod_{k=1}^d a_k(x_k) \quad (\mu^*_{\text{obs}}\text{-a.e.}).
 \end{equation}
 That is, $\mu_{\text{tech, obs}}$ already captures the exact higher-order copula interactions among the observable dimensions $(X_1, \dots, X_d)$.
 \item For $\mu^*_{\text{obs}}$-almost every $x_{\text{obs}} \in \Omega_{\text{obs}}$, the conditional probability kernel of the true economy on $\Omega_{\text{hid}}$ coincides with the conditional technological prior
 \begin{equation}
  \mu^*(dx_{\text{hid}} \mid x_{\text{obs}}) = \mu_{\text{tech}}(dx_{\text{hid}} \mid x_{\text{obs}}).
 \end{equation}
 In particular, under existence of a measurable mapping $f: \Omega_{\text{obs}} \to \Omega_{\text{hid}}$ such that $X_{\text{hid}} = f(X_{\text{obs}})$ $\mu^*$-a.e. (corresponding on discrete partitions to zero conditional Shannon entropy $H(X_{\text{hid}} \mid X_{\text{obs}}) = 0$ \cite{ElementsOfInformaionTheory}, or for square-integrable random variables to zero conditional variance $\text{Var}(X_{\text{hid}} \mid X_{\text{obs}}) = 0$ a.e.) - exact recovery on $\Omega_{\text{hid}}$ holds if and only if the prior conditional kernel is concentrated on the deterministic graph of $f$, i.e., $\mu_{\text{tech}}(dx_{\text{hid}} \mid x_{\text{obs}}) = \delta_{f(x_{\text{obs}})}(dx_{\text{hid}})$.
\end{enumerate}
\end{Theorem}
\begin{Proof}
By the chain rule for Kullback-Leibler divergence \cite{KLEntropy, ElementsOfInformaionTheory}, for every measure $\nu \in \Gamma$ with $\nu \ll \mu_{\text{tech}}$,
\begin{equation}
 \text{KL}(\nu \parallel \mu_{\text{tech}}) = \text{KL}(\nu_{\text{obs}} \parallel \mu_{\text{tech, obs}}) + \int_{\Omega_{\text{obs}}} \text{KL}\big(\nu(\cdot \mid x_{\text{obs}}) \parallel \mu_{\text{tech}}(\cdot \mid x_{\text{obs}})\big) \, d\nu_{\text{obs}}(x_{\text{obs}}).
 \label{Eq_KLChainRuleProof}
\end{equation}
Because the coordinate projections $\pi_k: \Omega \to \Omega_k$ ($k = 1, \dots, d$) depend exclusively on $x_{\text{obs}}$, the constraint $\nu \in \Gamma$ constrains only the observable marginal $\nu_{\text{obs}} \in \Gamma_{\text{obs}}$ and imposes no restriction on the conditional kernel $\nu(\cdot \mid x_{\text{obs}})$.

Consequently, minimizing (\ref{Eq_KLChainRuleProof}) over $\nu \in \Gamma$ decouples into two independent steps:
\begin{enumerate}
 \item For any fixed observable marginal $\nu_{\text{obs}} \in \Gamma_{\text{obs}}$, the integral of conditional KL divergences in (\ref{Eq_KLChainRuleProof}) is non-negative and attains its minimum value of $0$ if and only if
 \begin{equation}
  \hat{\mu}^*(dx_{\text{hid}} \mid x_{\text{obs}}) = \mu_{\text{tech}}(dx_{\text{hid}} \mid x_{\text{obs}}) \quad \text{for } \hat{\mu}^*_{\text{obs}}\text{-a.e. } x_{\text{obs}} \in \Omega_{\text{obs}}.
 \end{equation}
 \item The remaining term $\text{KL}(\nu_{\text{obs}} \parallel \mu_{\text{tech, obs}})$ is minimized over $\nu_{\text{obs}} \in \Gamma_{\text{obs}}$ by the unique observable $I$-projection $\hat{\mu}^*_{\text{obs}} = \arg\min_{\eta \in \Gamma_{\text{obs}}} \text{KL}(\eta \parallel \mu_{\text{tech, obs}})$. By Theorem \ref{Th_ReconstructionKL} and Csisz\'ar's theorem \cite{KLEntropy}, $\hat{\mu}^*_{\text{obs}} = \mu^*_{\text{obs}}$ holds if and only if $\frac{d\mu^*_{\text{obs}}}{d\mu_{\text{tech, obs}}}(x_1, \dots, x_d) = \prod_{k=1}^d a_k(x_k)$ $\mu^*_{\text{obs}}$-a.e.
\end{enumerate}
Combining both steps, $\hat{\mu}^* = \mu^*$ on $\Omega = \Omega_{\text{obs}} \times \Omega_{\text{hid}}$ if and only if $\hat{\mu}^*_{\text{obs}} = \mu^*_{\text{obs}}$ (Condition 1) and $\mu^*(dx_{\text{hid}} \mid x_{\text{obs}}) = \hat{\mu}^*(dx_{\text{hid}} \mid x_{\text{obs}}) = \mu_{\text{tech}}(dx_{\text{hid}} \mid x_{\text{obs}})$ $\mu^*_{\text{obs}}$-a.e. (Condition 2). Under Technological Rigidity ($X_{\text{hid}} = f(X_{\text{obs}})$ $\mu^*$-a.e.), the true conditional kernel is the Dirac measure $\mu^*(dx_{\text{hid}} \mid x_{\text{obs}}) = \delta_{f(x_{\text{obs}})}(dx_{\text{hid}})$, which has finite (zero) conditional KL divergence and is recovered by $\hat{\mu}^*$ if and only if $\mu_{\text{tech}}(dx_{\text{hid}} \mid x_{\text{obs}}) = \delta_{f(x_{\text{obs}})}(dx_{\text{hid}})$ $\mu^*_{\text{obs}}$-a.e.
\end{Proof}
\begin{Theorem}[The Information Pythagorean Theorem (Theorem 2.2 in \cite{KLEntropy})]
\label{Th_InformationPythagoreanTheorem}
Let $\mathcal{P}(\Omega)$ be the space of probability measures on $\Omega$, let $\mu_{\text{tech}} \in \mathcal{P}(\Omega)$ be a reference prior measure, and let $\Gamma \subset \mathcal{P}(\Omega)$ be a closed convex set of probability measures such that $\hat{\mu}^* = \arg\min_{\nu \in \Gamma} \text{KL}(\nu \parallel \mu_{\text{tech}})$ exists with $\text{KL}(\hat{\mu}^* \parallel \mu_{\text{tech}}) < \infty$. Then for any true measure $\mu^* \in \Gamma$ with $\text{KL}(\mu^* \parallel \mu_{\text{tech}}) < \infty$, the following inequality holds
\begin{equation}
 \text{KL}(\mu^* \parallel \mu_{\text{tech}}) \ge \text{KL}(\mu^* \parallel \hat{\mu}^*) + \text{KL}(\hat{\mu}^* \parallel \mu_{\text{tech}}).
 \label{Eq_InformationtriangleInequality}
\end{equation}
Furthermore, if $\Gamma = \Gamma(\mu_1, \dots, \mu_d)$ is defined by linear marginal constraints (with $\mu^* \ll \hat{\mu}^*$), the relationship holds with strict equality (see equation (3.1) in \cite{KLEntropy})
\begin{equation}
 \text{KL}(\mu^* \parallel \mu_{\text{tech}}) = \text{KL}(\mu^* \parallel \hat{\mu}^*) + \text{KL}(\hat{\mu}^* \parallel \mu_{\text{tech}}).
\end{equation}
The three terms can be given economic interpretation:
\begin{itemize}
 \item $\text{KL}(\mu^* \parallel \mu_{\text{tech}})$ - total information divergence between the true economy $\mu^*$ and the government's baseline technological prior $\mu_{\text{tech}}$.
 \item $\text{KL}(\hat{\mu}^* \parallel \mu_{\text{tech}})$ - the divergence accounted for by updating $\mu_{\text{tech}}$ to satisfy the observed institutional marginals $\mu_1, \dots, \mu_d$. Alternatively, this represents the information captured by the estimator under the assumed technological baseline.
 \item $\text{KL}(\mu^* \parallel \hat{\mu}^*)$ - the residual discrepancy between the true economy $\mu^*$ and the $I$-projection $\hat{\mu}^*$, caused by unmeasured cross-dimensional copula shifts or hidden-dimension deviations from $\mu_{\text{tech}}$.
\end{itemize}
\end{Theorem}
\begin{Proof}
Following \cite{KLEntropy}, for $t \in [0,1]$ define the convex mixture
\begin{equation}
 \nu_t = (1-t)\hat{\mu}^* + t\mu^* = \hat{\mu}^* + t(\mu^* - \hat{\mu}^*).
\end{equation}
Since $\Gamma$ is convex and $\hat{\mu}^*, \mu^* \in \Gamma$, we have $\nu_t \in \Gamma$ for all $t \in [0,1]$. Because $\hat{\mu}^*$ minimizes $\text{KL}(\nu \parallel \mu_{\text{tech}})$ over $\Gamma$, the directional derivative of $\phi(t) := \text{KL}(\nu_t \parallel \mu_{\text{tech}})$ at $t = 0^+$ must be non-negative:
\begin{equation}
 \left. \frac{d}{dt} \text{KL}(\nu_t \parallel \mu_{\text{tech}}) \right\vert_{t=0^+} \ge 0.
\end{equation}
Let $p^* = \frac{d\mu^*}{d\mu_{\text{tech}}}$ and $\hat{p} = \frac{d\hat{\mu}^*}{d\mu_{\text{tech}}}$ be the densities of $\mu^*$ and $\hat{\mu}^*$ with respect to $\mu_{\text{tech}}$, so $\nu_t$ has density $p_t = \hat{p} + t(p^* - \hat{p})$. Differentiating $\phi(t) = \int_\Omega p_t \log p_t \, d\mu_{\text{tech}}$ with respect to $t$ gives
\begin{equation}
 \phi'(t) = \int_\Omega (p^* - \hat{p}) \log p_t \, d\mu_{\text{tech}} + \int_\Omega (p^* - \hat{p}) \, d\mu_{\text{tech}}.
\end{equation}
Since $\mu^*$ and $\hat{\mu}^*$ are both probability measures, $\int_\Omega (p^* - \hat{p}) \, d\mu_{\text{tech}} = 1 - 1 = 0$. Evaluating at $t = 0^+$ yields the optimality inequality
\begin{equation}
 \int_\Omega p^* \log \hat{p} \, d\mu_{\text{tech}} \ge \int_\Omega \hat{p} \log \hat{p} \, d\mu_{\text{tech}} = \text{KL}(\hat{\mu}^* \parallel \mu_{\text{tech}}).
 \label{Eq_PythOptimality}
\end{equation}
Now compute the difference
\begin{align}
 \text{KL}(\mu^* \parallel \mu_{\text{tech}}) - \text{KL}(\mu^* \parallel \hat{\mu}^*)
 &= \int_\Omega p^* \log p^* \, d\mu_{\text{tech}} - \int_\Omega p^* \log\left(\frac{p^*}{\hat{p}}\right) d\mu_{\text{tech}} \notag \\
 &= \int_\Omega p^* \log \hat{p} \, d\mu_{\text{tech}}.
\end{align}
Applying (\ref{Eq_PythOptimality}) immediately yields (\ref{Eq_InformationtriangleInequality}):
\begin{equation}
 \text{KL}(\mu^* \parallel \mu_{\text{tech}}) - \text{KL}(\mu^* \parallel \hat{\mu}^*) \ge \text{KL}(\hat{\mu}^* \parallel \mu_{\text{tech}}).
\end{equation}
When $\Gamma = \Gamma(\mu_1, \dots, \mu_d)$ is defined by linear marginal constraints and $\log \hat{p}(x) = \sum_{k=1}^d \lambda_k(x_k) - 1$ as in Theorem \ref{Th_ReconstructionKL}, both $\mu^*$ and $\hat{\mu}^*$ share identical marginals $\mu_k$ on each $\Omega_k$, so
\begin{equation}
 \int_\Omega (p^* - \hat{p}) \log \hat{p} \, d\mu_{\text{tech}} = \sum_{k=1}^d \left(\int_{\Omega_k} \lambda_k(x_k) \, d\mu_k(x_k) - \int_{\Omega_k} \lambda_k(x_k) \, d\mu_k(x_k)\right) = 0,
\end{equation}
which turns (\ref{Eq_PythOptimality}) and (\ref{Eq_InformationtriangleInequality}) into strict equalities.
\end{Proof}
\begin{Corollary}
Since $I$-projection preserves the higher-order copula interactions and hidden-dimension conditional laws of the prior $\mu_{\text{tech}}$, any newly invented tax evasion scheme that alters the copula $C$ without changing the 1D marginals $\mu_k$ is invisible to the marginal constraints. Its entire divergence is absorbed into the unrecoverable error $\text{KL}(\mu^* \parallel \hat{\mu}^*)$. Furthermore, if $\mu^*$ assigns positive probability to a region where $\mu_{\text{tech}}$ vanishes ($\mu^* \not\ll \mu_{\text{tech}}$), then $\text{KL}(\mu^* \parallel \mu_{\text{tech}}) = +\infty$ and the prior must be recalibrated via ground-truth audits.
\end{Corollary}

\section{Compressed Sensing regime}
\label{Section_CS}

\begin{Theorem}[Exact Recovery of Sparse Shadow Economy (Theorem 1.2 in  \cite{CandesTaoLemma})]
\label{Th_CSRecovery}
Let $s^* \in \mathbb{R}^N$ be the unknown shadow-economy gap vector observed through $M \ll N$ aggregated checks $y = \Phi s^*$, where $\Phi \in \mathbb{R}^{M \times N}$. Assume:
\begin{enumerate}
 \item $K$-Sparsity in the Innovation Domain: The vector $s^*$ has at most $K$ non-zero entries ($K < M \ll N$):
 \begin{equation}
  \Vert s^* \Vert_0 = |\{j \in \{1, \dots, N\} \mid s^*_j \ne 0\}| \le K,
 \end{equation}
 where $\text{supp}(s^*) = \{j \mid s^*_j \ne 0\}$.
 \item Restricted Isometry Property (RIP) of Order $2K$: The sensing matrix $\Phi$ satisfies the RIP of order $2K$, meaning that for every $2K$-sparse vector $v \in \mathbb{R}^N$,
 \begin{equation}
  (1 - \delta_{2K}) \Vert v \Vert_2^2 \le \Vert \Phi v \Vert_2^2 \le (1 + \delta_{2K}) \Vert v \Vert_2^2,
 \end{equation}
 with restricted isometry constant strictly satisfying
 \begin{equation}
  \delta_{2K} < \sqrt{2} - 1.
 \end{equation}
\end{enumerate}
Then the convex $\ell_1$-minimization (Basis Pursuit) problem \cite{DataDrivenScience}
\begin{equation}
 \hat{s} = \arg\min_{s \in \mathbb{R}^N} \Vert s \Vert_1 \quad \text{subject to} \quad \Phi s = y
\end{equation}
has a unique solution, and recovers the true shadow economy vector exactly
\begin{equation}
 \hat{s} = s^*.
\end{equation}
\end{Theorem}
\begin{Proof}
We adapt the geometric null-space and shelling proof of Cand\`es \cite{CandesTaoLemma}. Let $\hat{s}$ be any minimizer of the $\ell_1$ problem, and define the reconstruction error vector $h := \hat{s} - s^*$. Since $\Phi \hat{s} = y = \Phi s^*$, we have $h \in \ker(\Phi)$.

Let $T_0 = \text{supp}(s^*)$, so $|T_0| \le K$. By optimality of $\hat{s}$, $\Vert \hat{s} \Vert_1 = \Vert s^* + h \Vert_1 \le \Vert s^* \Vert_1$. Decomposing the $\ell_1$ norm over $T_0$ and its complement $T_0^c$ and applying the reverse triangle inequality gives
\begin{equation}
 \Vert s^* \Vert_1 \ge \Vert s^*_{T_0} + h_{T_0} \Vert_1 + \Vert h_{T_0^c} \Vert_1 \ge \Vert s^*_{T_0} \Vert_1 - \Vert h_{T_0} \Vert_1 + \Vert h_{T_0^c} \Vert_1.
\end{equation}
Since $\Vert s^*_{T_0} \Vert_1 = \Vert s^* \Vert_1$, it yields the $\ell_1$ cone condition
\begin{equation}
 \Vert h_{T_0^c} \Vert_1 \le \Vert h_{T_0} \Vert_1.
 \label{Eq_CSConeTrap}
\end{equation}
Economic interpretation is as follows: Equation (\ref{Eq_CSConeTrap}) represents a geometric trap: any evasion perturbation $h \in \ker(\Phi)$ that attempts to hide within the null space of the state audit matrix $\Phi$ without increasing the $\ell_1$ norm must concentrate at least half of its $\ell_1$ mass on the true $K$-sparse support $T_0$.

In the next step, we partition $T_0^c$ into disjoint subsets $T_1, T_2, \dots$ of size $K$ in decreasing order of the magnitudes $|h_i|$ (so $|h_i| \ge |h_m|$ for all $i \in T_{j-1}$, $m \in T_j$, $j \ge 2$). For every $j \ge 2$ and $m \in T_j$,
\begin{equation}
 |h_m| \le \frac{1}{K}\sum_{i \in T_{j-1}} |h_i| = \frac{1}{K}\Vert h_{T_{j-1}} \Vert_1,
\end{equation}
which implies $\Vert h_{T_j} \Vert_2 \le K^{-1/2} \Vert h_{T_{j-1}} \Vert_1$. Summing over $j \ge 2$, using (\ref{Eq_CSConeTrap}) and the Cauchy-Schwarz ($\Vert h_{T_0} \Vert_1 \le \sqrt{K}\Vert h_{T_0} \Vert_2$) gives, so-called Shell Inequality,
\begin{equation}
 \sum_{j \ge 2} \Vert h_{T_j} \Vert_2 \le K^{-1/2} \sum_{j \ge 1} \Vert h_{T_j} \Vert_1 = K^{-1/2} \Vert h_{T_0^c} \Vert_1 \le K^{-1/2} \Vert h_{T_0} \Vert_1 \le \Vert h_{T_0} \Vert_2.
 \label{Eq_ShellingInequality}
\end{equation}
Since $h \in \ker(\Phi)$, we have $\Phi h_{T_0 \cup T_1} = -\Phi h_{(T_0 \cup T_1)^c} = -\sum_{j \ge 2} \Phi h_{T_j}$. By the RIP of order $2K$ and Lemma 2.1 of Cand\`es \cite{CandesTaoLemma} (which states $|\langle \Phi u, \Phi v \rangle| \le \delta_{2K} \Vert u \Vert_2 \Vert v \Vert_2$ for any two vectors $u, v$ supported on disjoint sets of cardinality at most $K$), we obtain
\begin{align}
 (1 - \delta_{2K}) \Vert h_{T_0 \cup T_1} \Vert_2^2
 &\le \Vert \Phi h_{T_0 \cup T_1} \Vert_2^2 = \left\langle \Phi h_{T_0 \cup T_1}, -\sum_{j \ge 2} \Phi h_{T_j} \right\rangle \notag \\
 &= -\sum_{j \ge 2} \langle \Phi h_{T_0}, \Phi h_{T_j} \rangle - \sum_{j \ge 2} \langle \Phi h_{T_1}, \Phi h_{T_j} \rangle \notag \\
 &\le \delta_{2K} \left(\Vert h_{T_0} \Vert_2 + \Vert h_{T_1} \Vert_2\right) \sum_{j \ge 2} \Vert h_{T_j} \Vert_2 \notag \\
 &\le \sqrt{2}\delta_{2K} \Vert h_{T_0 \cup T_1} \Vert_2 \sum_{j \ge 2} \Vert h_{T_j} \Vert_2,
\end{align}
where the last inequality uses $\Vert h_{T_0} \Vert_2 + \Vert h_{T_1} \Vert_2 \le \sqrt{2(\Vert h_{T_0} \Vert_2^2 + \Vert h_{T_1} \Vert_2^2)} = \sqrt{2}\Vert h_{T_0 \cup T_1} \Vert_2$. Dividing by $(1 - \delta_{2K})\Vert h_{T_0 \cup T_1} \Vert_2$ and substituting (\ref{Eq_ShellingInequality}) yields
\begin{equation}
 \Vert h_{T_0} \Vert_2 \le \Vert h_{T_0 \cup T_1} \Vert_2 \le C(\delta_{2K}) \frac{\Vert h_{T_0^c} \Vert_1}{\sqrt{K}} \le C(\delta_{2K}) \frac{\Vert h_{T_0} \Vert_1}{\sqrt{K}} \le C(\delta_{2K}) \Vert h_{T_0} \Vert_2,
\end{equation}
where
\begin{equation}
 C(\delta_{2K}) = \frac{\sqrt{2}\delta_{2K}}{1 - \delta_{2K}}.
\end{equation}
Since $\delta_{2K} < \sqrt{2} - 1$ implies $C(\delta_{2K}) < 1$, this inequality forces $\Vert h_{T_0} \Vert_2 = 0$, so $h_{T_0} = 0$. By (\ref{Eq_CSConeTrap}), $\Vert h_{T_0^c} \Vert_1 \le \Vert h_{T_0} \Vert_1 = 0$, so $h = 0$ and $\hat{s} = s^*$, as claimed.
\end{Proof}

\section*{Acknowledgments}
RK work was partially funded by the EU Horizon Europe MSCA grant No 101300007.

During the preparation of this work, the authors used generative AI tools to assist with language polishing and structuring the presentation of proofs. After using these tools, the authors independently verified all mathematical derivations and take full responsibility for the content of the publication.

\appendix